\documentclass[12pt]{article}
\usepackage[margin=1in]{geometry}

\usepackage[utf8]{inputenc}
\usepackage{color,xcolor,graphicx,amsmath,float, amssymb,mathtools,url, amsfonts,amssymb,physics,mathbbol,float,tikz,caption,subcaption,multicol,multirow,authblk,hyperref,cleveref}
\usepackage[subnum]{cases}

\usepackage{eufrak}
\def\ss{\mathfrak{a}_{\rm scale}}

\hypersetup{
  colorlinks=true,
   linkcolor=blue!80!black,
  urlcolor=blue!60!black
  ,citecolor=blue!80!black
  }

\date{\vspace{-5ex}}
\begin{document}
\title{\vspace{-2.0cm}\bf Near-inflection point inflation and production of dark matter during reheating }

\author{
\small{
Anish Ghoshal\\\vspace{-0.4cm}Institute of Theoretical Physics, Faculty of Physics, University of Warsaw,\\ ul. Pasteura 5, 02-093 Warsaw, Poland,\\ 
email: anish.ghoshal@fuw.edu.pl\\Gaetano Lambiase\\Dipartimento di Fisica ”E.R. Caianiello”, Universita’ di Salerno, I-84084 Fisciano (Sa), Italy\\ Gruppo Collegato di Salerno, I-84084 Fisciano (Sa), Italy\\
email:lambiase@sa.infn.it
\\Supratik Pal\\Physics and Applied Mathematics Unit, Indian Statistical Institute, Kolkata-700108, India\\ Technology Innovation Hub on Data Science, Big Data Analytics and Data Curation,
Indian Statistical Institute, Kolkata-700108, India\\
email:supratik@isical.ac.in
\\
Arnab Paul\\Physics and Applied Mathematics Unit, Indian Statistical Institute, Kolkata-700108, Indi\\School of Physical Sciences, Indian Association for the Cultivation of Science, Kolkata-700032, India \\
email:arnabpaul9292@gmail.com\\
Shiladitya Porey\thanks{corresponding author}\\Department of Physics and Astronomy, Novosibirsk State University,\\
email: shiladityamailbox@gmail.com
}
}

\maketitle
\vspace{-1.0cm}
\begin{abstract}
\textit{
We study slow roll single field inflationary scenario and the production of non-thermal fermionic dark matter, together with standard model Higgs, during reheating. For the inflationary scenario, we have considered two models of polynomial potential – one is symmetric about the origin and another one is not. We fix the coefficients of the potential from the current Cosmic Microwave Background ({CMB})~data from \textit{Planck}/\textsc{Bicep}. Next, we explore the allowed parameter space on the coupling $(y_\chi)$ with inflaton and mass $(m_\chi)$ of dark matter ({DM}) particles $(\chi)$ produced during reheating and satisfying {CMB}~and several other cosmological constraints.
}
\end{abstract}

\section{Introduction}
Cosmic inflation which is postulated as a fleeting cosmological epoch, occurred at the very early time of the universe. During this primordial epoch, spacetime expanded exponentially resulting in statistical homogeneity and isotropy on large angular scales, the exceedingly flat universe, and providing a proper explanation for the horizon
problem. In addition to that, inflation can generate quantum fluctuations, which transform into scalar and tensor perturbations. Scalar perturbation acts as the mechanism for the formation of the large-scale structure, while tensor perturbation is responsible for generating gravitational wave. The simplest way to fabricate such an epoch is to assume that the universe was dominated by the energy density of a single scalar field, called inflaton, minimally coupled to gravity and having canonical kinetic energy, slowly rolling along the slope of the potential. However, current data from {CMB}~measurements, e.g. \textit{Planck}~\cite{Aghanim:2018eyx} and \textsc{Bicep}~\cite{BICEPKeck:2022mhb}, favour plateau-like potential over the inflaton-potential of the form $V(\phi)\propto \phi^p$ with $p\geq 1$. One of the other alternatives to get such a potential is to consider inflection-point inflation. 

On the other hand, {CMB}~measurements  suggest that approximately one-quarter of the total mass-energy density of the present universe is in the form of Dark Matter ({DM}) whose true nature is still not known with certainty. All proposed possible particles of {DM}~can be categorized into two groups - Weakly Interacting Massive Particles ({WIMP}) and Feebly Interacting Massive Particles ({FIMP}). Till now, the signature of the presence of {WIMP}~particles has not been detected in particle detector experiments~\cite{LZ:2022ufs}. In that case, {FIMP}~which were never in thermal equilibrium with the relativistic plasma of the universe, seems more favorable as the viable {DM}~candidate~\cite{Billard:2021uyg}.

In the paper \cite{Ghoshal:2022jeo} we studied a single unified model of inflation and the production of non-thermal dark matter particles. For the inflationary part, we have considered two small-field inflection point inflationary scenarios. We have also assumed direct coupling between the inflaton and the {DM}, a vector-like fermionic field $\chi$ which transforms as gauge singlet under the {SM}~gauge groups. The inflaton either decays to {DM}~or may undergo scattering with the dark sector to produce the observed relic. As we will see, additional irreducible gravitational interaction may also mediate the {DM}~production, either by 2-to-2 annihilation of the Standard Model ({SM}) Higgs bosons or of the inflatons during the reheating era.

This paper is organized as follows: in Section~\ref{Sec:Inflection-point Inflation Models}, we discuss the condition of getting an inflection point for a single field potential. In Section~\ref{sec:Inflection-point achieved with Linear term}, we study the slow roll inflationary scenario for two potentials and find the location of inflection point and fix the coefficients of the potentials from {CMB}~data. Reheating and production of dark matter have been discussed in Section~\ref{Sec:Reheating}. Section~\ref{Sec:Conclusions and Discussion} contains conclusion.

\section{Inflection-point inflation models}
\label{Sec:Inflection-point Inflation Models}
%
%
Near the location of the inflection point, the potential takes a plateau-like shape. Because of that, inflection point of the inflationary potential is important for the slow roll inflationary scenario.
If inflaton starts rolling along the potential from the vicinity of the inflection point, the number of e-foldings (described in Section~\ref{sec:Inflection-point achieved with Linear term}) increases without significant change in the inflaton value. %

To determine the stationary inflection point of an inflationary potential ${\cal V}(\psi)$ of a single scalar field $\psi$, we need the solution of 
\begin{eqnarray} \label{eq:condition_for_inflection_point}
\frac{\text{d} {\cal V}}{\text{d} \psi} = \frac{\text{d}^2 {\cal V}}{\text{d} \psi^2}=0   \,.
\end{eqnarray} 
In the following sections (Section~\ref{sec:Inflection-point achieved with Linear term}) we discuss two different slow roll small-field inflationary scenarios, where each of the inflationary potentials possesses an inflection point. 
\section{Slow roll inflationary scenario}
\label{sec:Inflection-point achieved with Linear term}
The Lagrangian density we are interested in, is given by in $\hbar=c=k_{B}=1$ unit, 
\begin{align}
\label{Eq:Lagrangian density-ModI}
\mathcal{L}_{I} &= \frac{M_P^2}{2} \mathcal{R}
+\mathcal{L}_{KE, INF} + U_{INF} 
+ {\cal L}_{KE, \chi} 
  - U_\chi(\chi) 
  + {\cal L}_{KE, H} 
  - U_H(H)
 + \mathcal{L}_{reh} 
 \,,
\end{align}
where $M_P\simeq 2.4 \times 10^{18}\, \text{GeV}$ is the reduced Planck mass and ${\cal R}$
is the Ricci scalar with metric-signature $(+,-,-,-)$. $\mathcal{L}_{KE, INF}$ and $U_{INF}$ are respectively the kinetic energy and potential energy term of the single scalar inflaton. Since, those two terms are function of inflaton, they alter when we change the model of inflation. In this work, we use $\Phi$ to symbolize inflaton for \text{Model}  I inflation and $\varphi$ for \text{Model} II. Accordingly, 
\begin{numcases}{U_{INF} \equiv}
    U_\Phi=  V_0 + a \, \Phi - b \, \Phi^2 + d \, \Phi^4 \qquad  \text{(for Model I)}  \,, \label{eq:inflation potential of model I}
    \\
    U_\varphi= p \, \varphi^2 - q \, \varphi^4 + \text{w} \, \varphi^6 \qquad  \text{(for Model II)}  \,.  \label{eq:inflation potential of model II}
\end{numcases}
Here $V_0$, $a$, $b$, $d$, $p, q$, and $\text{w}$ are all assumed to be positive, real; and we choose $d, \text{w} > 0$. The potential of Eq.~\eqref{eq:inflation potential of model I} contains a term of linear order of inflaton. Due to this term $ U_\Phi$ is not symmetric about the origin. On the contrary, the $ U_\varphi$ is symmetric about the origin. 
In Eq.~\eqref{Eq:Lagrangian density-ModI},
${\cal L}_{KE, \chi}$, and ${\cal L}_{KE, H}$ represent the kinetic energy of the vector-like fermionic {DM}, $\chi$, and Standard Model ({SM})~Higgs field, $H$, respectively. And the potential term for $\chi$ and $H$ are given by - 
\begin{eqnarray}
&&U_\chi(\chi) =  m_{\chi} \bar{\chi}\chi  \, ,\\
&& U_H(H) = -m_H^2 H^\dagger H  + \lambda_H \left( H^\dagger H \right)^2 \,.
\end{eqnarray}
Furthermore, the last term on the right side of Eq.~\eqref{Eq:Lagrangian density-ModI}, $\mathcal{L}_{reh} $, takes care of the interactions of $\chi$ and $H$ with $\Phi (\varphi)$ during reheating 
and it is defined as
\begin{numcases}{\mathcal{L}_{reh} \equiv}
\mathcal{L}_{reh,I} = 
- y_\chi \Phi \bar{\chi}\chi - \lambda_{12} \Phi H^\dagger H - \lambda_{22} \Phi^2 H^\dagger H  \qquad  \text{(for Model I)} \,,\label{eq:reheating lagrangian for modelI}\\
\mathcal{L}_{reh,II} =- y_\chi \varphi \bar{\chi}\chi - \lambda_{12} \varphi H^\dagger H - \lambda_{22} \varphi^2 H^\dagger H  \qquad  \text{(for Model II)}  \,, \label{eq:reheating lagrangian for modelII}
\end{numcases}

where $\lambda_{12}$, $\lambda_{22}$, and Yukawa-like $y_\chi$ are the couplings of {SM}~Higgs and fermionic {DM}~with inflaton.

During the slow roll inflationary epoch, contribution from the terms except the first three terms in Eq.~\eqref{Eq:Lagrangian density-ModI} is negligible. 
The slow-roll condition is measured in terms of 
four potential-slow-roll parameters – $\epsilon_V,\eta_V,\xi_V,$ and $\sigma_V$. 
During slow roll inflationary epoch, $\left|\epsilon_V\right|,\left|\eta_V\right|,\left|\xi_V\right|,\left|\sigma_V\right| \ll 1$.  These four potential-slow-roll parameters for \text{Model} I are defined as 
\begin{eqnarray}
\epsilon_V &\approx& \frac{M_P^2}{2}\left( \frac{U_\Phi^\prime}{U_\Phi}\right)^2
=M_P^2\frac{\left(a-2 b \, \Phi +4 d \, \Phi ^3\right)^2}{2 \left(\Phi  \left(a-b \, \Phi +d \,  \Phi^3\right)+V_0\right){}^2}   \,,
\label{eq:epsilonV}
\\
\eta_V &\approx& M_P^2\frac{U_\Phi^{\prime \prime}}{U_\Phi} 
= -M_P^2\frac{2 \left(b-6 d \, \Phi ^2\right)}{\Phi  \left(a-b \, \Phi +d \, \Phi ^3\right)+V_0}   \,,
\label{eq:etaV}\\
%
%
\xi_V&
\approx& M_P^4\frac{U_\Phi^\prime U_\Phi^{\prime \prime \prime}}{U_\Phi^2} 
= M_P^4\frac{24 d \, \Phi  \left(a-2 b \, \Phi +4 d \, \Phi ^3\right)}{\left(\Phi  \left(a-b \, \Phi +d \, \Phi^3\right)+V_0\right){}^2}  \,,
\label{eq:xiV}
\\
\sigma_V &\approx& M_P^6\frac{{U_\Phi^\prime}^2 U_\Phi^{\prime \prime \prime \prime}}{U_\Phi^3} = M_P^6\frac{24 d \, \left(a-2 b \, \Phi +4 d \, \Phi ^3\right)^2}{\left(\Phi  \left(a-b \, \Phi +d \, \Phi^3\right)+V_0\right){}^3}  \,.
 \end{eqnarray}
Here, prime denotes derivative with respect to inflaton. For \text{Model} II inflation, the potential-slow-roll parameters are 
\begin{eqnarray}
&&\epsilon_V= M_P^2\frac{2 \left(p \varphi -2 q \varphi ^3+3 \text{w} \varphi ^5\right)^2}{\left(p \varphi ^2-q \varphi ^4+\text{w} \varphi ^6\right){}^2} \,, \\
&&\eta_V= M_P^2\frac{2 \left(p-6 q \varphi ^2+15 \text{w} \varphi ^4\right)}{p \varphi ^2-q \varphi ^4+\text{w} \varphi^6}  \,, \\
&&\xi_V =M_P^4\frac{48 \varphi ^2 \left(-q+5 \text{w} \varphi ^2\right) \left(p-2 q \varphi ^2+3 \text{w} \varphi  ^4\right)}{\left(p \varphi ^2-q \varphi ^4+\text{w} \varphi ^6\right){}^2}  \,,\\
&&\sigma_V=M_P^6\frac{96 \left(-q+15 \text{w} \varphi ^2\right) \left(p \varphi -2 q \varphi ^3+3 \text{w} \varphi
   ^5\right)^2}{\left(p \varphi ^2-q \varphi ^4+\text{w} \varphi ^6\right){}^3}   \,.
\end{eqnarray} 

By the time 
 any one of these slow-roll parameters 
becomes $\sim 1$ at $\Phi \sim \Phi_{\rm end}$ \text{(for Model I)}~or at  $\varphi \sim \varphi_{\rm end}$ \text{(for Model II)}, slow roll inflation terminates. The duration of slow roll inflation is measured in terms of the total number of e-foldings, $\mathcal{N}_{\rm CMB, \, tot}$ as
\begin{equation}\label{eq:def-e-fold}
\mathcal{N}_{\rm CMB, \, tot}=M_P^{-2}\int_{\Phi_{\rm end}(\varphi_{\rm end})}^{\Phi_{\rm CMB} (\varphi_{\rm CMB})}
\frac{U_{INF}}{U'_{INF}}\, \text{d} \Phi (\varphi)
=\int_{\Phi_{\rm end}(\varphi_{\rm end})}^{\Phi_{\rm CMB}(\varphi_{\rm CMB})} \frac{1}{\sqrt{2 \epsilon_V}} \, \text{d} \Phi   (\varphi)\,, 
\end{equation}
where $\Phi_{\rm CMB} (\varphi_{\rm CMB})$ 
is the inflaton value at which the length scale, which had previously left the causal horizon during inflation, has reentered during the period of recombination.

Moreover, inflation generates primordial scalar and tensor perturbations. 
The primordial scalar and tensor power spectrum for '$k$'-th Fourier mode are defined as
\begin{eqnarray}
&& \mathcal{P}_s \left( k \right) = A_s \left(  \frac{k}{k_*} \right)^{n_s -1 + (1/2) \alpha_s \ln(k/k_*) + (1/6)\beta_s (\ln(k/k_*))^2 }  \label{eq:define scalar power spectrum}\,, \\
&& \mathcal{P}_h \left( k \right) = A_t \left(  \frac{k}{k_*} \right)^{n_t + (1/2) d n_t/d \ln k \ln(k/k_*) + \cdots } \,,\label{eq:define tensor power spectrum}
\end{eqnarray}
where $k_*=0.05 \text{Mpc}^{-1}$; $n_s$ and $n_t$ are the scalar and tensor spectral index, $\alpha_s$ is the running of scalar spectral index, and $\beta_s$ is called the 'running of running'. 
Moreover, in Eq.~\eqref{eq:define scalar power spectrum}-\eqref{eq:define tensor power spectrum}, $A_s$ and $A_t$ are the normalizations. The relation between $A_s$ and inflationary potential is 
\begin{eqnarray}\label{eq:As}
A_s \approx \frac{U_{INF}}{24 \pi^2  M_P^4\, \epsilon_V} \approx \frac{2 U_{INF}}{3 \pi^2 M_P^4 \, r} \,.
\end{eqnarray} 
Here, $r$ is the tensor-to-scalar ratio. 
$r$, $n_s$, 
$\alpha_s$ and 
$\beta_s$ depend on potential-slow-roll parameters as 
\begin{align}
& r = \frac{A_t}{A_s}\approx 16 \epsilon_V  \,.\label{eq:r aprrox EtaV} \quad \quad
n_s = \frac{\text{d} \ln \mathcal{P}_s }{\text{d} \ln k} 
= 1+ 2\eta_V - 6\epsilon_V \,,  \\
& \alpha_s \equiv\frac{\text{d} n_s}{\text{d} \ln k}  =16 \epsilon_V \eta_V -24 \epsilon_V^2 - 2\xi_V \,. & \\
& \beta_s \equiv 
\frac{\text{d}^2 n_s}{\text{d} \ln k^2}   
= 
-192 \epsilon_V^3 + 192 \epsilon_V^2 \eta_V - 32 \epsilon_V \eta_V^2 -24 \epsilon_V \xi_V  + 2\eta_V \xi_V +2 \sigma_V \,. 
\end{align}

The observed values of all these inflation parameters  measured at $\Phi=\Phi_{\rm CMB}$ (at $k_{*}\simeq 0.05 \text{Mpc}^{-1}$) from \textit{Planck}, \textsl{WMAP},~%
and other {CMB}~observations are presented in Table~\ref{Table:PlanckData}.
\footnote{  T and E corresponds to temperature and E-mode polarisation of CMB.
}
\begin{table}[ht]
\begin{center}
\caption{ \it {CMB}~constraints on inflationary parameters.} \label{Table:PlanckData}
\begin{tabular}{ |c| c| c|c| }
\hline
$\ln(10^{10} A_s)$ & $3.047\pm 0.014$ & $68\%$, TT,TE,EE+lowE+lensing+BAO & 
\cite{Aghanim:2018eyx}  \\
 \hline
 $n_s$ & $0.9647\pm 0.0043$ & $68\%$, TT,TE,EE+lowE+lensing+BAO & 
 \cite{Aghanim:2018eyx} \\ 
 \hline 
 $\text{d} n_s/\text{d} \ln k$ & $0.0011\pm0.0099$ & $68\%$, TT,TE,EE+lowE+lensing+BAO &  
 \cite{Aghanim:2018eyx}  \\  
 \hline
 $\text{d}^2 n_s/\text{d} \ln k^2$  & $0.009 \pm  0.012$ & $68\%$, TT,TE,EE+lowE+lensing+BAO &  
 \cite{Aghanim:2018eyx} \\
 \hline 
 $r$ & $0.014^{+0.010}_{-0.011}\, \text{and}$ &  $ 95 \%  \,, \text{BK18, \textsc{Bicep}3, \textit{Keck Array}~2020,}$& \cite{Aghanim:2018eyx, BICEPKeck:2022mhb,BICEP:2021xfz,Campeti:2022vom}\\
   & $ <0.036 $ & and \textsl{WMAP} and \textit{Planck}~CMB polarization &  \\
 \hline 
\end{tabular}
\end{center}
\end{table}%
\subsection{Estimating coefficients from  {CMB}~%
data}\label{sec:Planck Data}
In this subsection, we find the location of inflection points and also, fix the coefficient of the potentials of both inflationary models, mentioned in Eq.~\eqref{eq:inflation potential of model I} and Eq.~\eqref{eq:inflation potential of model II}, from the CMB~data. At first, we start the calculation with \text{Model}~I. 
Solution of Eq.~\eqref{eq:condition_for_inflection_point} provides the location of inflection point \text{for Model I}~potential 
\begin{equation}\label{eq:inflection point}
\Phi_0 = \frac{3 a}{4 b}  \qquad \text{ when }\, d= \frac{ 8 b^3}{27 a^2} \,.
\end{equation}
%
%
To fix the coefficients of the potential of Eq.~\eqref{eq:inflation potential of model I},
following~\cite{Hotchkiss:2011gz,Chatterjee:2014hna}, 
we can write
\begin{eqnarray}
\begin{pmatrix}
	\Phi_{\rm CMB} & \Phi_{\rm CMB}^2 & \Phi_{\rm CMB}^4\\
	1 & 2\Phi_{\rm CMB} & 4\Phi_{\rm CMB}^3 \\
	0  & 2     & 12\Phi_{\rm CMB}^2 
\end{pmatrix}
\begin{pmatrix}
	a \\
	b\\
	d 
\end{pmatrix}
&&= \begin{pmatrix}
	U_\Phi (\Phi_{\rm CMB}) - V_0\\
	U'_\Phi(\Phi_{\rm CMB})\\
	U''_\Phi(\Phi_{\rm CMB}) 
\end{pmatrix} \,,
\end{eqnarray}
where $d$ is known from Eq.~\eqref{eq:inflection point} and $U_\Phi(\Phi_{\rm CMB}), U'_\Phi(\Phi_{\rm CMB})$ and $U''_\Phi(\Phi_{\rm CMB})$ can be derived using Eq.~\eqref{eq:epsilonV}, \eqref{eq:etaV},
\eqref{eq:xiV}, \eqref{eq:As}, 
\eqref{eq:r aprrox EtaV}
as
\begin{eqnarray}
&&U_\Phi(\Phi_{\rm CMB}) = \frac{3}{2}A_s r \pi^2  M_P^4\, , \label{eq:U}\\
&&U'_\Phi(\Phi_{\rm CMB}) = \frac{3}{2}\sqrt{\frac{r}{8}}\left(A_s r \pi^2 \right)  M_P^3\,, \label{eq:U'}\\
&&U''_\Phi(\Phi_{\rm CMB}) = \frac{3}{4}\left(\frac{3r}{8} + n_s -1\right)\left(A_s r \pi^2 \right)  M_P^2\,. \label{eq:U"}
\end{eqnarray}

\begin{figure}[!h]
\begin{center}
\begin{tabular}[c]{cc}
\begin{subfigure}[c]{0.45\textwidth}
 \includegraphics[width=\linewidth, height=0.6\linewidth ]{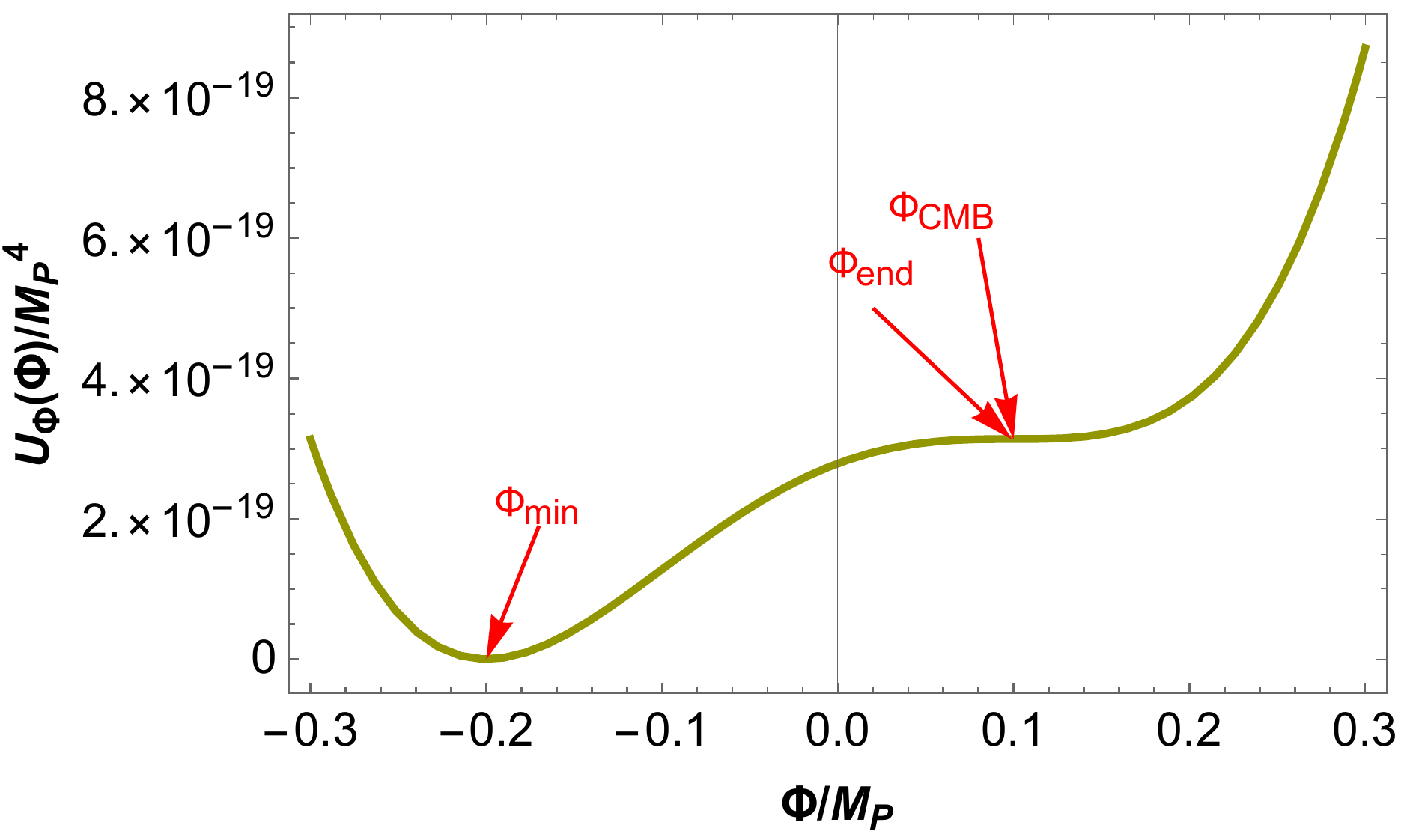}
  \label{fig:mesh1}
\end{subfigure}&%
\hspace{30pt}
\begin{subfigure}[c]{0.45\textwidth}
  \includegraphics[width=\linewidth, height=0.6\linewidth
  ]{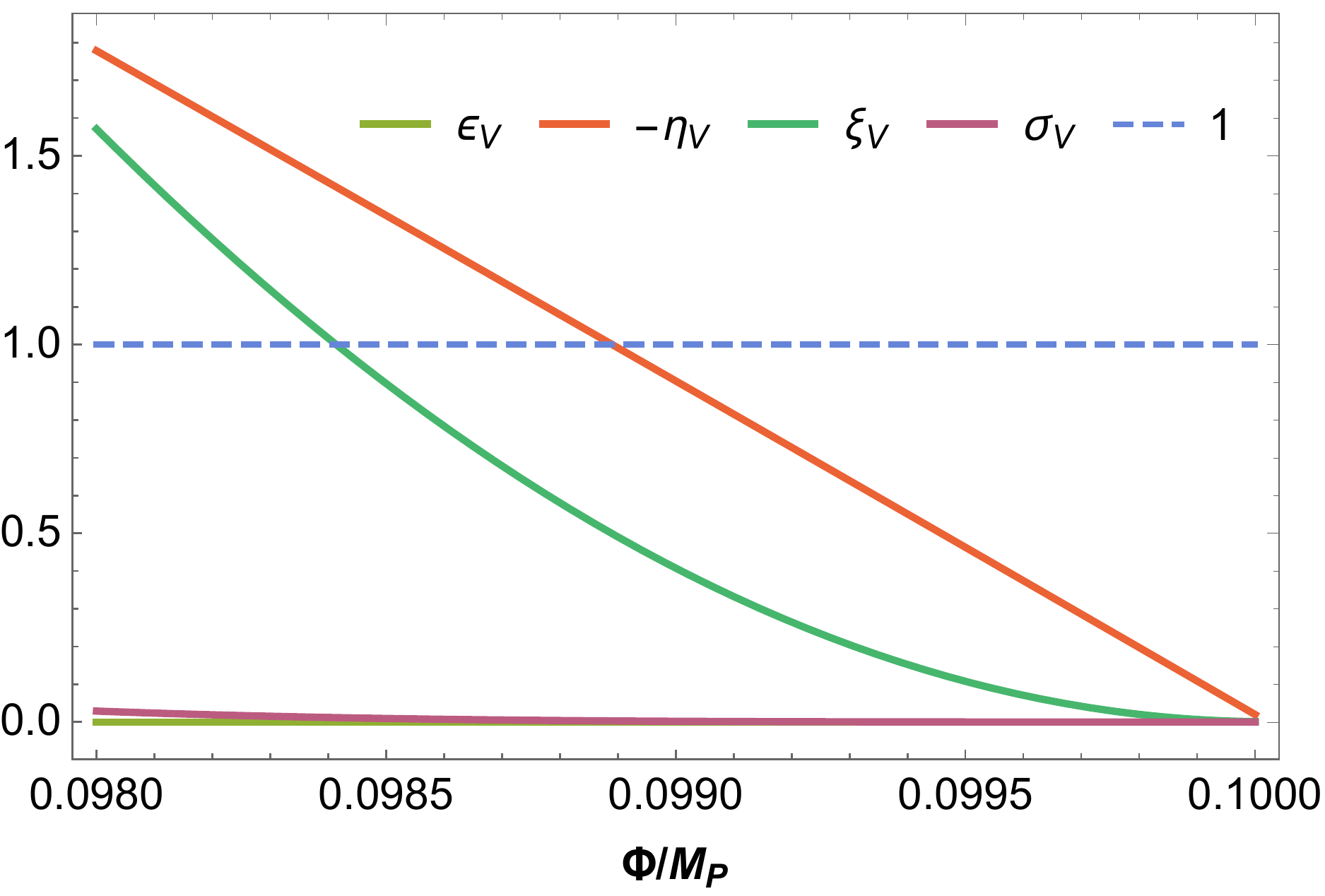}
  \label{fig:sub2}
\end{subfigure}\\
\begin{subfigure}[c]{0.45\textwidth}
 \includegraphics[width=\linewidth, height=0.6\linewidth]{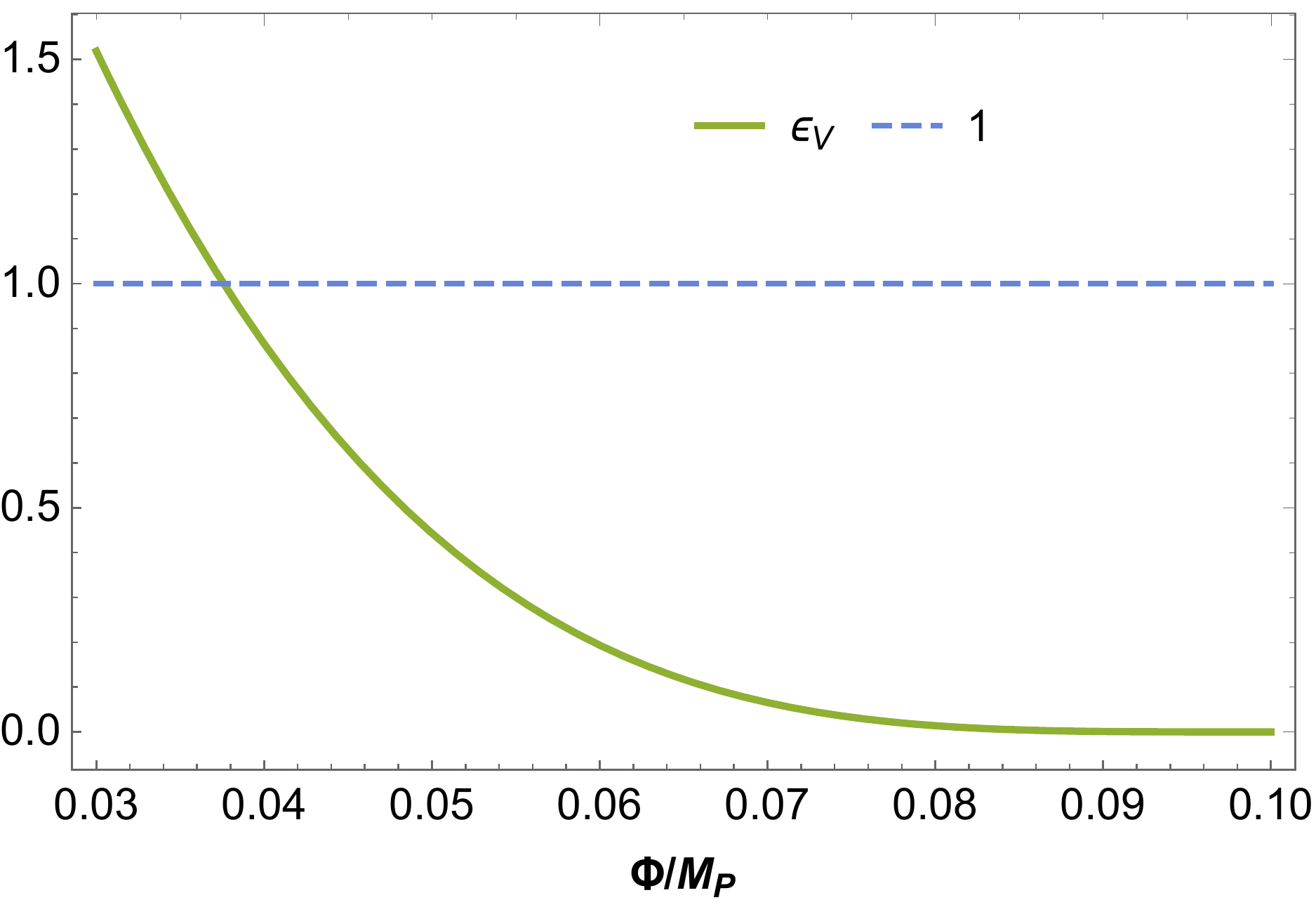}
\end{subfigure}&%
\hspace{30pt}
\begin{subfigure}[c]{.45\textwidth}
  \includegraphics[width=\linewidth, height=0.6\linewidth]{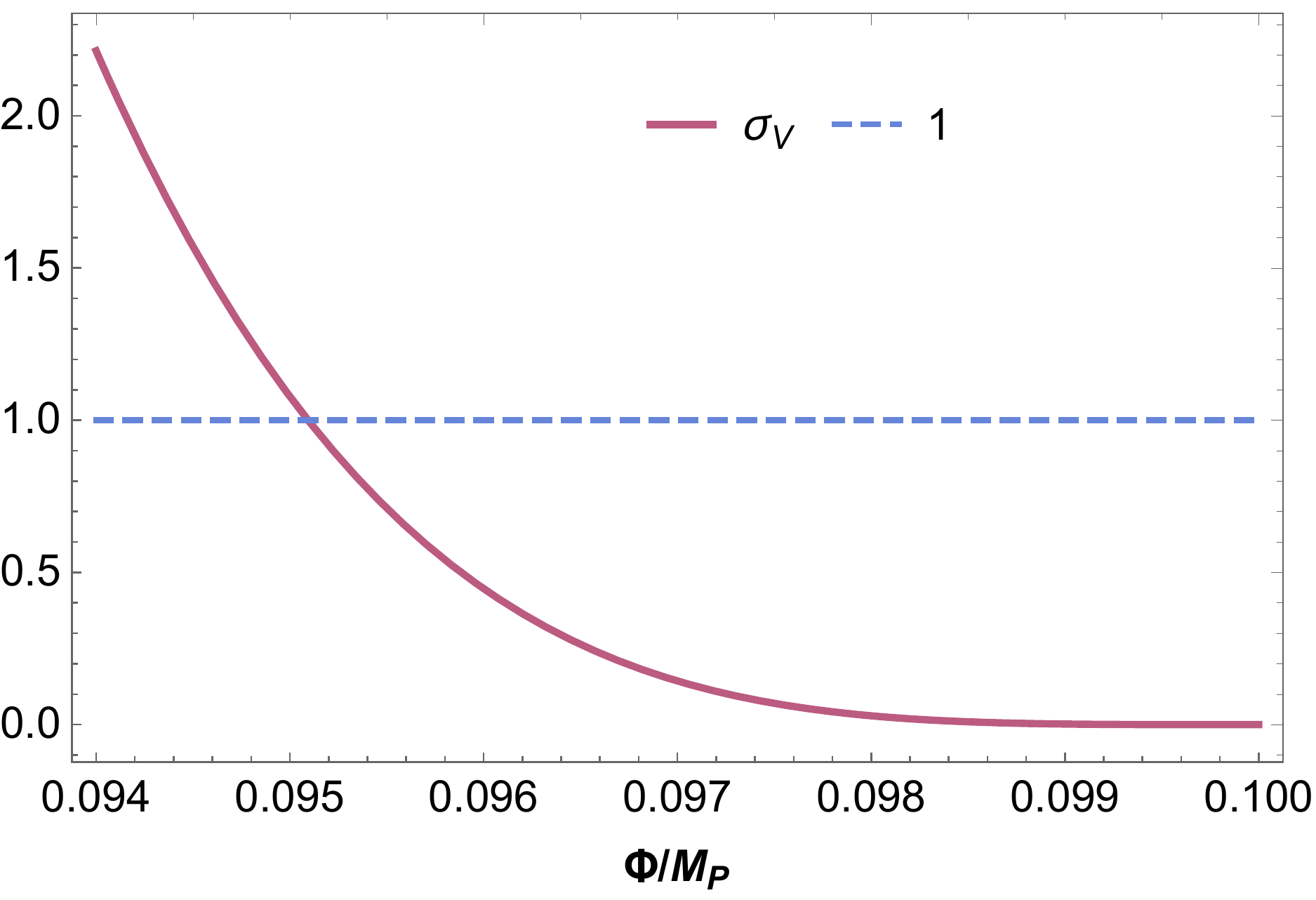}
   \label{fig:sub2}
\end{subfigure}\\
\end{tabular}    

\end{center}
\vspace{-20pt}
\caption{\it \raggedright%
In the top-left panel: normalised inflaton-potential of \text{Model} I inflation  as a function of '$\varphi/M_P$' for benchmark value shown in Table~\ref{Tab:Model I benchmark values}. The evolution of inflationary slow-roll parameters ($\epsilon_V,-\eta_V, \xi_V,\sigma_V$) as a function of $\Phi/M_P$ is presented in the top-right panel; second row - left panel: $\epsilon_V$, and second row – right panel: $\sigma_V$ of \text{Model} I slow roll inflation against $\Phi/M_P$ are shown individually for benchmark values listed in Table~\ref{Tab:Model I benchmark values}. 
The dashed line is for $1$. Whenever $\left|\eta_V\right|$ becomes $\sim 1$, the slow roll inflation ends. From these figures, it is clearly visible that $\left|\epsilon_V \right|<\left| \sigma_V\right|< \left|\xi_V \right|<\left|\eta_V\right|$ during the slow-roll regime.
}
\label{fig:potential_plot_linear_term_inflation}
\end{figure}

Using these together with Table~\ref{Table:PlanckData}, we can find the coefficients of the potential. However, for cosmological purpose, it is adequate to design the potential in a way such that $\Phi_{\rm CMB}$ is adjacent to $\Phi_0$~\cite{Garcia-Bellido:2017mdw}. In order to implement this, let us modify the potential (Eq.~\eqref{eq:inflation potential of model I}) as
\begin{eqnarray}\label{eq:modified potential of model I}
U_\Phi(\Phi) = V_0 + A \, \Phi - B \, \Phi^2 + d \, \Phi^4 \,,
\end{eqnarray} 
with $A= a (1-\beta^I_1), B= b(1-\beta^I_2)$ (where $\beta^I_1, \beta^I_2$ are dimensionless) 
and in the limit $\beta^I_1,\beta^I_2 \to 0$, the slope of the potential vanishes at $\Phi_{0}$. Using this modification, 
we have found the benchmark value for this potential which is exhibited in Table~\ref{Tab:Model I benchmark values}, and using this value, the evolution of the potential and slow roll parameters with $\Phi$ are illustrated in Fig.~\ref{fig:potential_plot_linear_term_inflation}. From this Fig.~\ref{fig:potential_plot_linear_term_inflation} it is clear that $\sigma_V,\xi_V,\epsilon_V<\left|\eta_V\right|$. Besides, at $\Phi=\Phi_{\rm CMB}$,  $\epsilon_V,\left|\eta_V\right|, \xi_V,\sigma_V <<1$, and at $\Phi=\Phi_{\rm end}$, $\left|\eta_V\right| \simeq  1$. 
This last condition leads to the ending of slow roll phase.


\begin{table}[H]
    \centering
        \caption{ \it Benchmark value for linear term potential (\text{Model} I) ($\Phi_{\rm min}$ is the minimum of potential in Eq.~\eqref{eq:modified potential of model I}) }
    \label{Tab:Model I benchmark values}
\vspace{-20pt}
\begin{center}
\begin{tabular}{ |c| c | c|c|c| c| }
\hline 
$V_0/M_P^4$ & $a/M_P^3$ & $b/M_P^2$ & $d$ & $\beta^I_1$ & $\beta^I_2$\\
 \hline 
 $2.788\times 10^{-19}$ &  $9.29\times 10^{-19}$   & $6.966\times 10^{-18}$  & $1.16\times 10^{-16}$   & $6 \times 10^{-7}$ & $6 \times 10^{-7}$ \\
 \hline 
\end{tabular}

\vspace{0.5pt}
\begin{tabular}{ |c| c |c|c| c |}
\hline
$\Phi_{\rm CMB}/M_P$ & $\Phi_{\rm end}/M_P$ & $\Phi_{\rm min}/M_P$         & $\Phi_{0}/M_P$\\
\hline 
 $0.1$ & $0.098889 $ & $-0.200045$    &      $0.100022$ \\
\hline
\end{tabular}

\vspace{0.5pt}
\begin{tabular}{ |c| c |c|c| c |c|}
\hline
$r$ & $n_s$ & $A_s$  & e-folding & $\alpha_s$ & $\beta_s$  \\
\hline 
$9.87606\times 10^{-12}$ & $0.960249$ & $2.10521\times 10^{-9} $ & $53.75$ & $-1.97\times 10^{-3}$ & $ -3.92\times10^{-5}$ \\
 \hline 
\end{tabular}
\end{center}
\end{table}


Next, we follow similar steps for the inflationary potential of \text{Model}~II. The potential of Eq.~\eqref{eq:inflation potential of model II} has an inflection point at
\begin{eqnarray}
\varphi_{0}= \frac{\sqrt{q}}{\sqrt{3\,  \text{w}}} \qquad \text{ for } \, p= \frac{ q^2}{3 \, \text{w}} \,.
\end{eqnarray} 


\begin{figure}[htp]
\centering
\begin{subfigure}{0.45\textwidth}
  \centering
 \includegraphics[width=\linewidth]{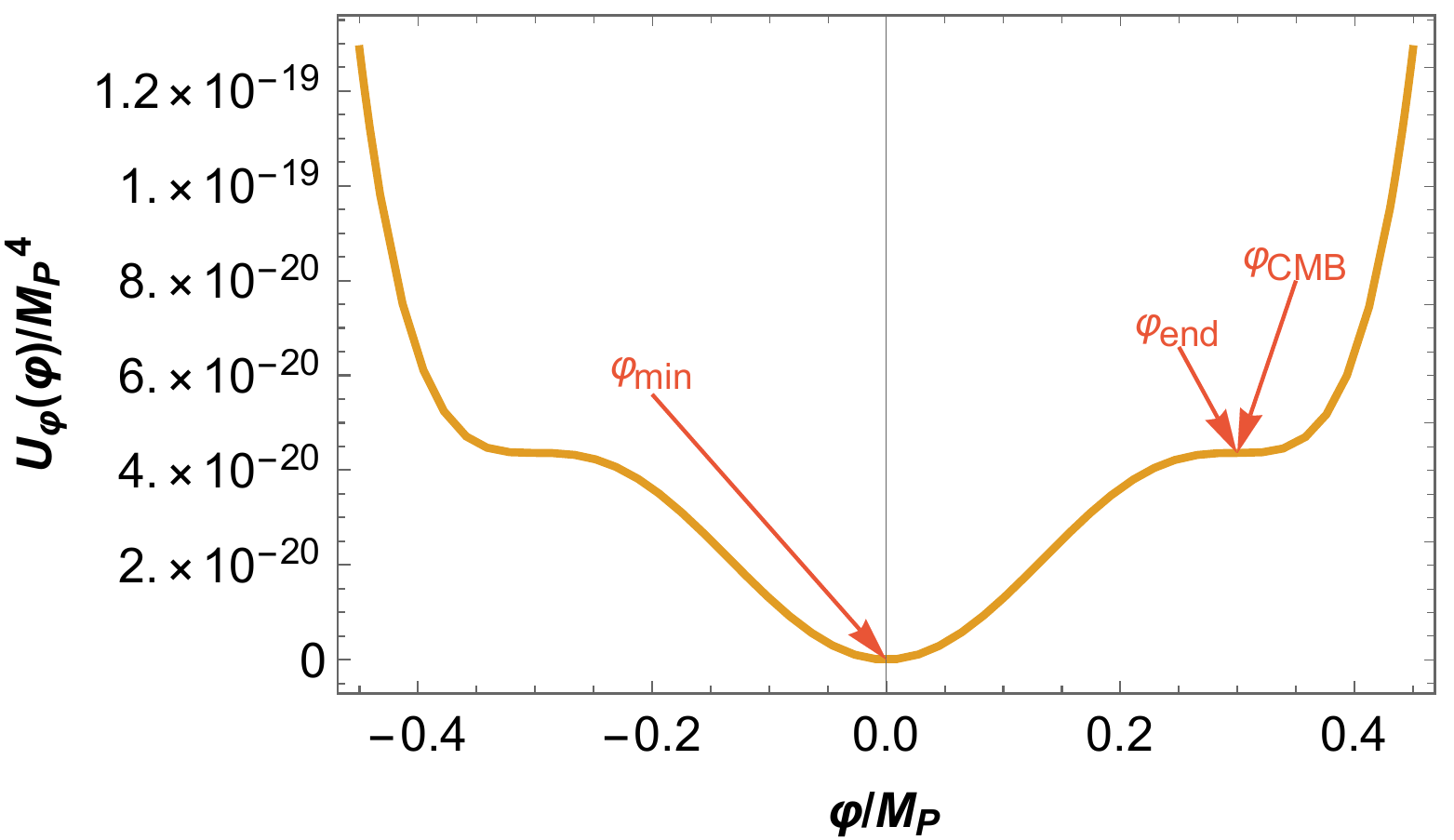}
  \label{fig:mesh1}
\end{subfigure}%
\hspace{30pt}
\begin{subfigure}{.45\textwidth}
  \centering
  \includegraphics[width=\linewidth]{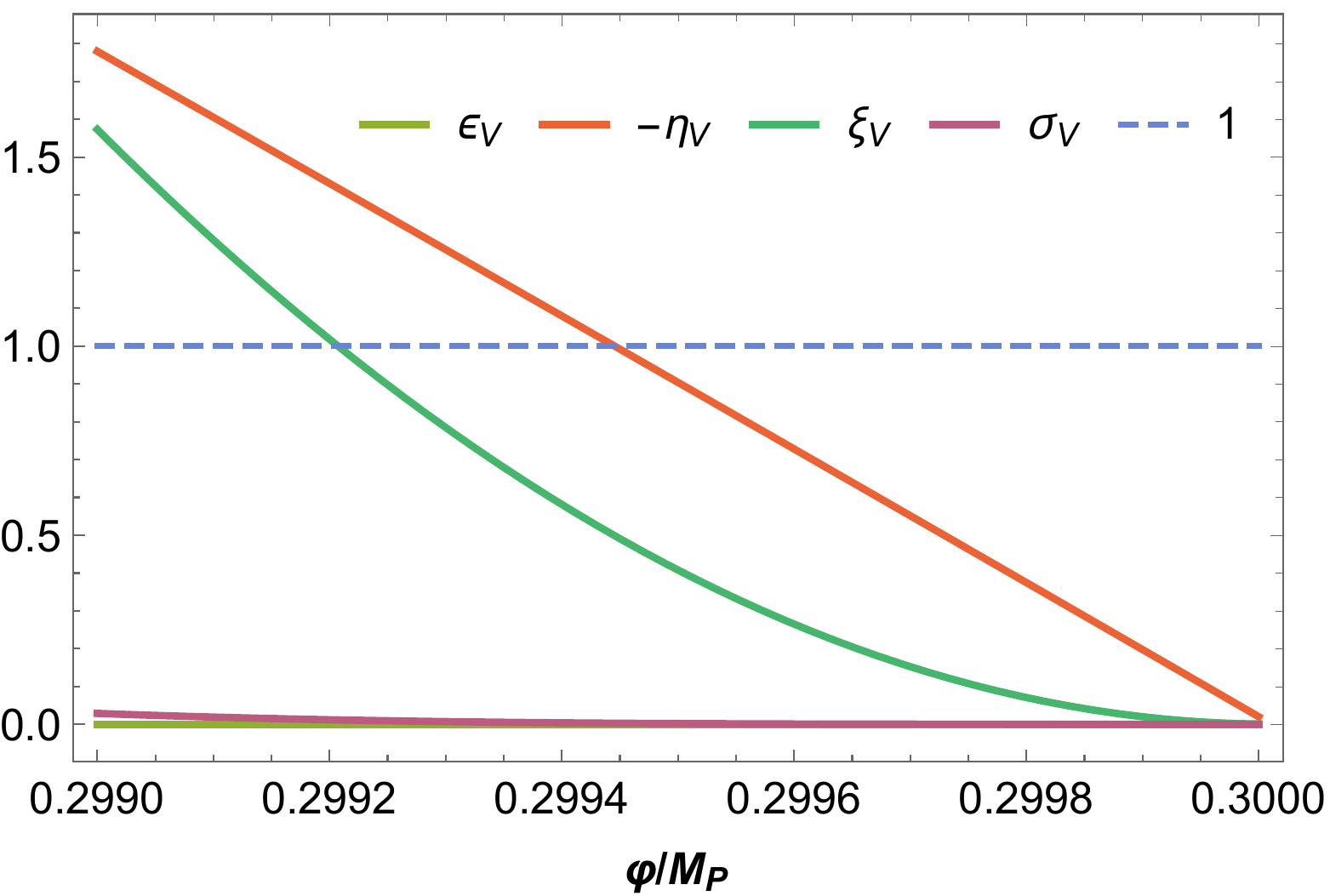}
  \label{fig:sub2}
\end{subfigure}
\begin{subfigure}{0.45\textwidth}
  \centering
 \includegraphics[width=\linewidth]{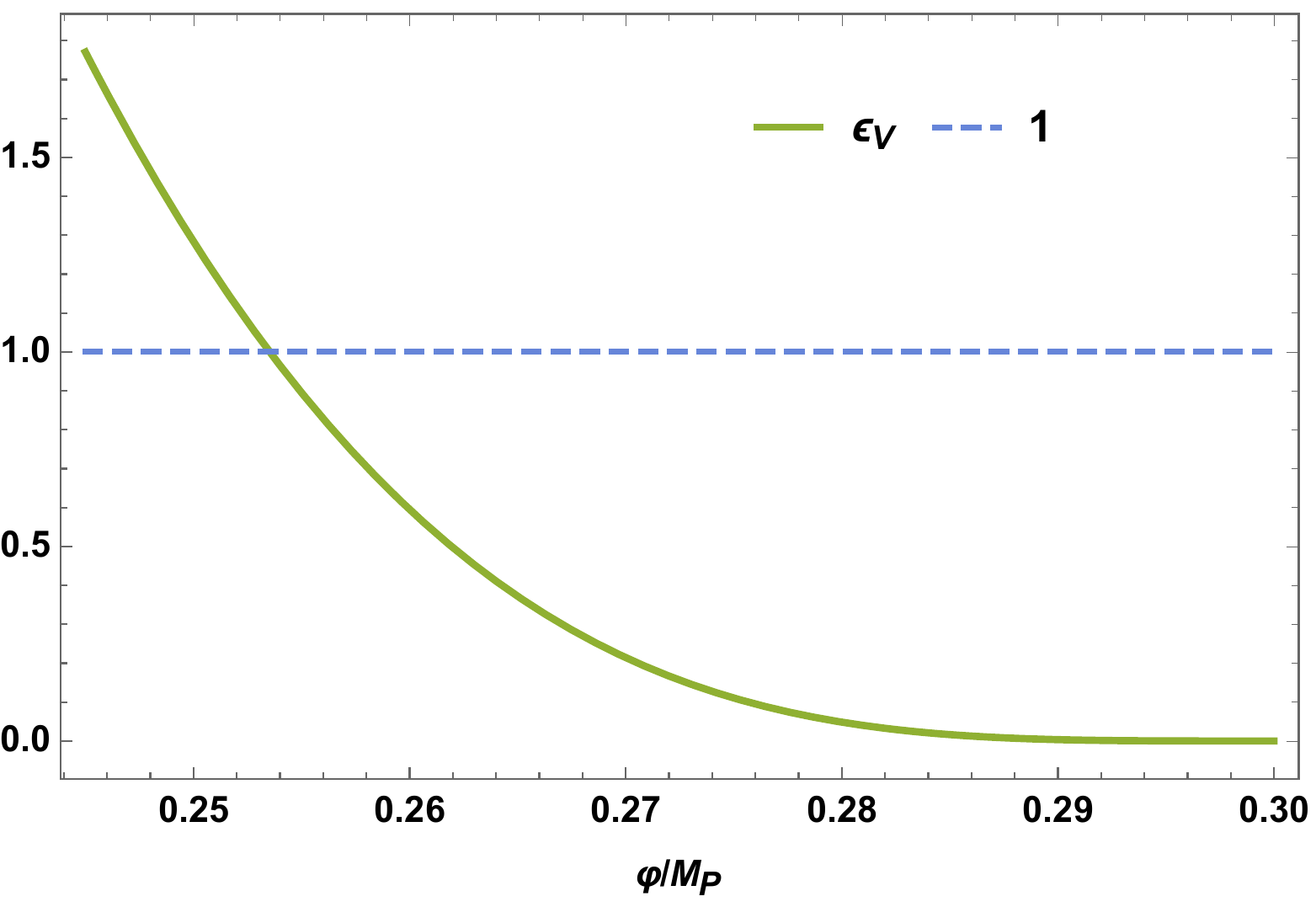}
   \label{fig:mesh1}
\end{subfigure}%
\hspace{30pt}
\begin{subfigure}{.45\textwidth}
  \centering
  \includegraphics[width=\linewidth]{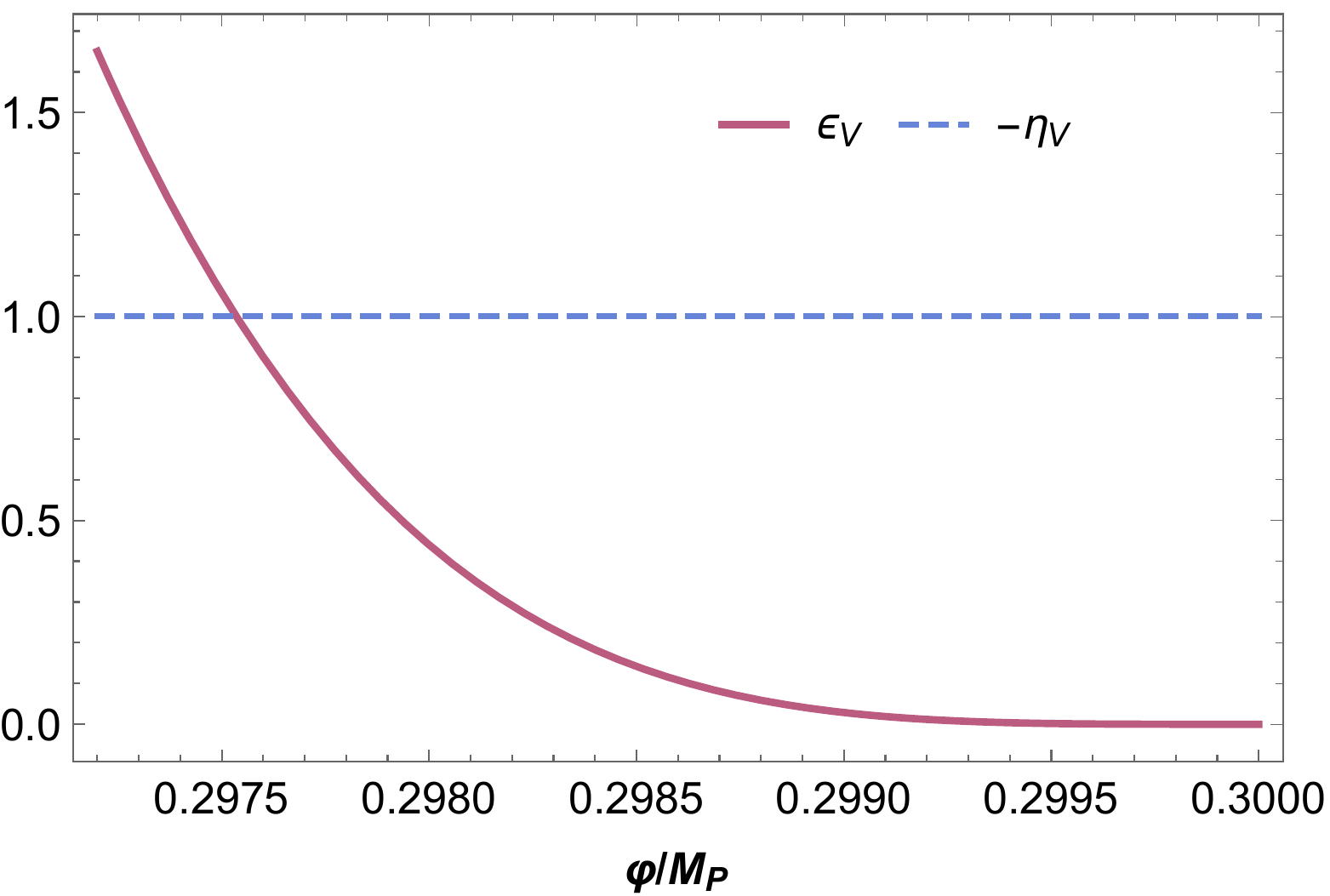}
   \label{fig:sub2}
\end{subfigure}
\vspace{-20pt}
\caption{\it \raggedright Top-left panel: evolution of normalised inflaton-potential of \text{Model} II for benchmark value from Table~\ref{Tab:Model II benchmark values}. Top-right panel: absolute values of four slow roll parameters ($\epsilon_V, -\eta_V,\xi_V,\sigma_V $) are plotted against $\varphi/M_P$.~Left and right panel of the second row displays $\epsilon_V$ and $\sigma_V$, respectively, against $\varphi/M_P$ for benchmark values mentioned in Table~\ref{Tab:Model II benchmark values}. 
The dashed line indicates $1$. These graphs demonstrate that $\left|\epsilon_V \right|<\left| \sigma_V\right|< \left|\xi_V \right|<\left|\eta_V\right|<1$ during the slow-roll inflation, similar to what we have found in \text{Model} I. 
}
\label{fig:potential_plot_sextic_inflation}
\end{figure}


Likewise, we can also redefine the potential \text{Model}~II as 
%
\begin{eqnarray}\label{eq:modified potential of model II}
U_\varphi(\varphi)  = 
p \, \varphi^2 - Q \, \varphi^4 + \text{W} \, \varphi^6 \,,
\end{eqnarray} 
such that $Q= q(1-\beta^{II}_1)$ and $\text{W}=w (1-\beta^{II}_2)$ and $\beta^{II}_1$, $\beta^{II}_2$ have zero mass dimension.
Then, we can estimate $p,q$ and $\text{w}$, and the values are mentioned in Table~\ref{Tab:Model II benchmark values}. For this value, the variation of $U_\varphi(\varphi)$ of Eq.~\eqref{eq:modified potential of model II} and $\epsilon_V,\left|\eta_V\right|, \xi_V,\sigma_V $ as a function of $\varphi$ is shown in Fig.~\ref{fig:potential_plot_sextic_inflation}.
The slow roll inflationary phase ends at $\varphi_{\rm end}$ when  $\left|\eta_V\right|\simeq 1$  (because for \text{Model} II $\epsilon_V<\left|\eta_V\right|$).


\begin{table}[H]
    \centering
        \caption{ \it Benchmark values for sextic potential ($\varphi_{\rm min}$ is the minimum of potential Eq.~\eqref{eq:modified potential of model II})}
    \label{Tab:Model II benchmark values}
\vspace{-20pt}
\begin{center}
\begin{tabular}{ | c |c|c| c |c|}
\hline 
 $p/M_P^2$ & $q$ & $\text{w} M_P^2$ & $\beta^{II}_1$     & $\beta^{II}_2$\\
 \hline 
  $1.45\times 10^{-18}$   & $1.62\times 10^{-17}$  & $5.98\times 10^{-17}$   &  $1.53\times 10^{-8}$   & $1.53\times 10^{-8}$ \\
 \hline 
\end{tabular}

\vspace{0.5pt}
\begin{tabular}{ |c| c |c|c| c |}
\hline
$\varphi_{\rm CMB}/M_P$ & $\varphi_{\rm end}/M_P$ & $\varphi_{\rm min}/M_P$         & $\varphi_{0}/M_P$\\
\hline 
$0.3$ & $0.299444$ & $ 0 $     &  $0.300011$\\
\hline
\end{tabular}

\vspace{0.5pt}
\begin{tabular}{ |c| c |c|c| c |c|  }
\hline
$r$ & $n_s$ & $A_s$  & e-folding & $\alpha_s$ & $\beta_s$       \\
\hline 
$1.4 \times 10^{-12}$ & $0.96001$ & $2.10521\times 10^{-9} $ & $60.247$ & $-1.487\times 10^{-3}$ & $ -2.972\times 10^{-5}$  \\
 \hline 
\end{tabular}
\end{center}
\end{table}

\section{Stability analysis}
\label{Sec:Stability analysis}
In this section, we  attempt to determine the upper bound of $y_\chi$ and $\lambda_{12}$ so that ${\cal L}_{reh,I}$ and ${\cal L}_{reh,II}$ do not affect the inflationary scenario set forth in Section~\ref{sec:Inflection-point achieved with Linear term}. 
The Coleman–Weinberg (CW) radiative correction at 1-loop order
to the inflaton-potential is given by~\cite{Drees:2021wgd} -
\begin{eqnarray}
V_{\rm CW}
=\sum_{j
} \frac{n_j}{64\pi^2} (-1)^{2s_j}\widetilde{m}_j^4
\left[ \ln\left( \frac{\widetilde{m}_j^2 
}{\mu^2} \right) - c_j \right]  \,.
\end{eqnarray}

Here, $j\equiv H, \chi$ and inflaton; $n_{H,\chi}=4$, $n_j$ for inflaton is  $1$. Furthermore, $s_H =0$, $s_\chi=1/2$, and $s_{\Phi(\varphi)}=0$. 
$\widetilde{m}_j$ is inflaton dependent mass of the component $j$ and $\mu$ is the renormalization scale, which is taken $\sim \Phi_{0}$ \text{(for Model I)} or $\varphi_{0}$ \text{(for Model II)}. 
Besides, $c_j=\frac{3}{2}$.  %
%
Now, the 
second derivative of the CW term w.r.t. inflaton is
\begin{align}
	%
	& V_{\rm CW}^{\prime\prime} =  \sum\limits_{j
	} \frac{n_j}{32 \pi^2} (-1)^{2 s_j} 
	\left\{ \left[ \left(\left(\widetilde{m}_j^{2}\right)^{\prime} \right)^2
	+ \widetilde{m}_j^2 \left(\widetilde{m}_j^{2}\right)^{\prime\prime} \right]
	\ln \left(\frac{\widetilde{m}_j^2}{\mu^2} \right)
	- \widetilde{m}_j^2 \left(\widetilde{m}_j^{2}\right)^{\prime\prime}  \right\}\,.   \label{eq:second derivative test}
\end{align}
%
In the next two subsections, we investigate the stability 
relative to the couplings $y_\chi$ and $\lambda_{12}$ for the two inflation-potentials
~(Eq.~\eqref{eq:modified potential of model I}) and Eq.~\eqref{eq:modified potential of model II}) we have considered.

\subsection{Stability analysis for linear term inflation}
\label{sec:Stability analysis for linear term inflation}
%

From Eq.~\eqref{eq:reheating lagrangian for modelI}, the field-depended mass of the $\chi$ and $H$  are respectively 
\begin{eqnarray}
& \widetilde{m}_{\chi}^2 (\Phi) = \left( m_\chi + y_\chi \Phi \right)^2  \,,\\
& \widetilde{m}_{H}^2 (\Phi) = m_H^2 + \lambda_{12} \Phi \,.
\end{eqnarray} 

For the stability of the inflation-potential, the terms of the order of $\lambda_{12}^2$ and $y_\chi^2$  on the right-hand side in Eq.~\eqref{eq:second derivative test} should be less than corresponding tree level terms from 
Eq.~\eqref{eq:modified potential of model I} -
\begin{align}
& V_{\rm tree}^{\prime\prime} (\Phi_{0}) \equiv U''_\Phi(\Phi_{0}) =  \frac{32 b^3 \Phi_{0}^2}{9 a^2}-2 b (1-\beta )
\,,\label{eq:tree-level-second-derivative-modelI}
\end{align}
where $\beta^I_1 = \beta^I_2=\beta^I$ (as we have chosen the benchmark value $\beta^I_1 = \beta^I_2$). 
The second derivative (Eq.~\eqref{eq:second derivative test}) of CW term for Higgs field is 
\begin{align}
& \left|V_{{\rm CW},H}^{\prime \prime}
\right|= \frac{\lambda_{12}^2 }{8 \pi ^2} \ln
\left(\frac{\lambda_{12} \Phi }{\Phi_0^2}\right) \, .\label{eq:upper limit of l12-coupling model I}
\end{align}

\begin{figure}[htp]
\centering
\begin{subfigure}{0.45\textwidth}
  \centering
 \includegraphics[width=\linewidth]{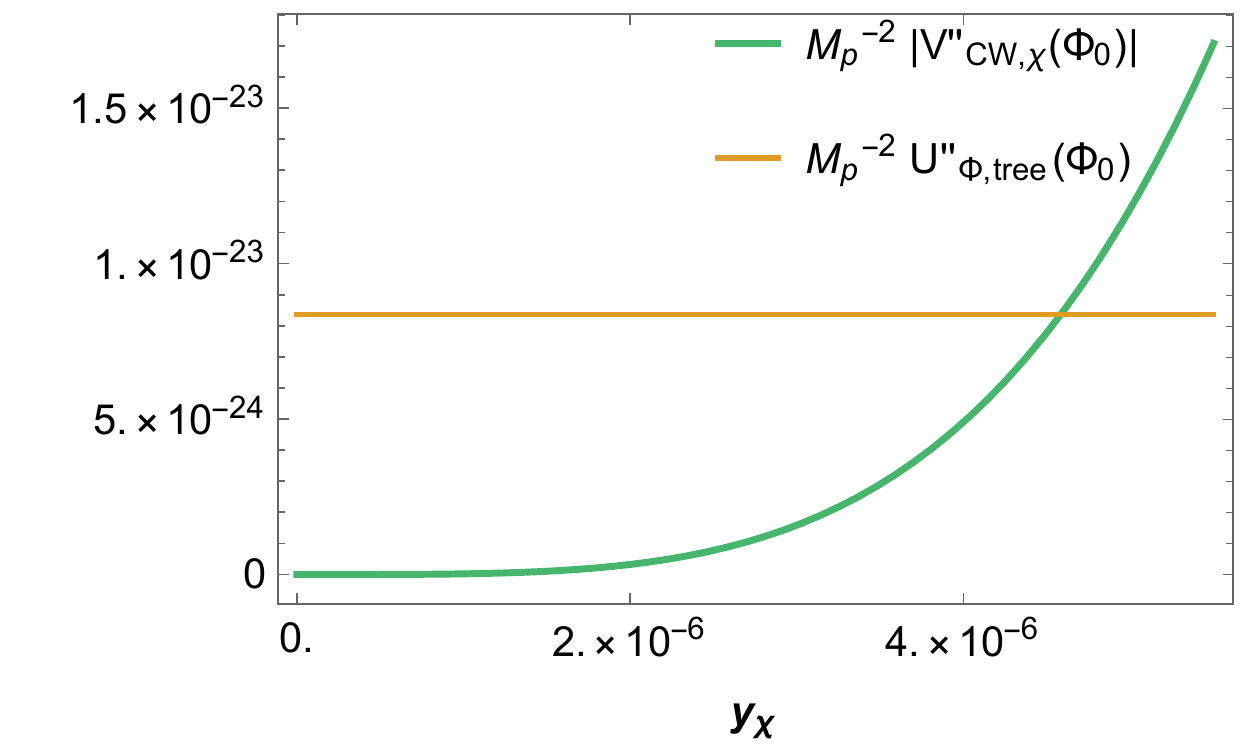}
\end{subfigure}%
\hspace{30pt}
\begin{subfigure}{.45\textwidth}
  \centering
  \includegraphics[width=\linewidth]{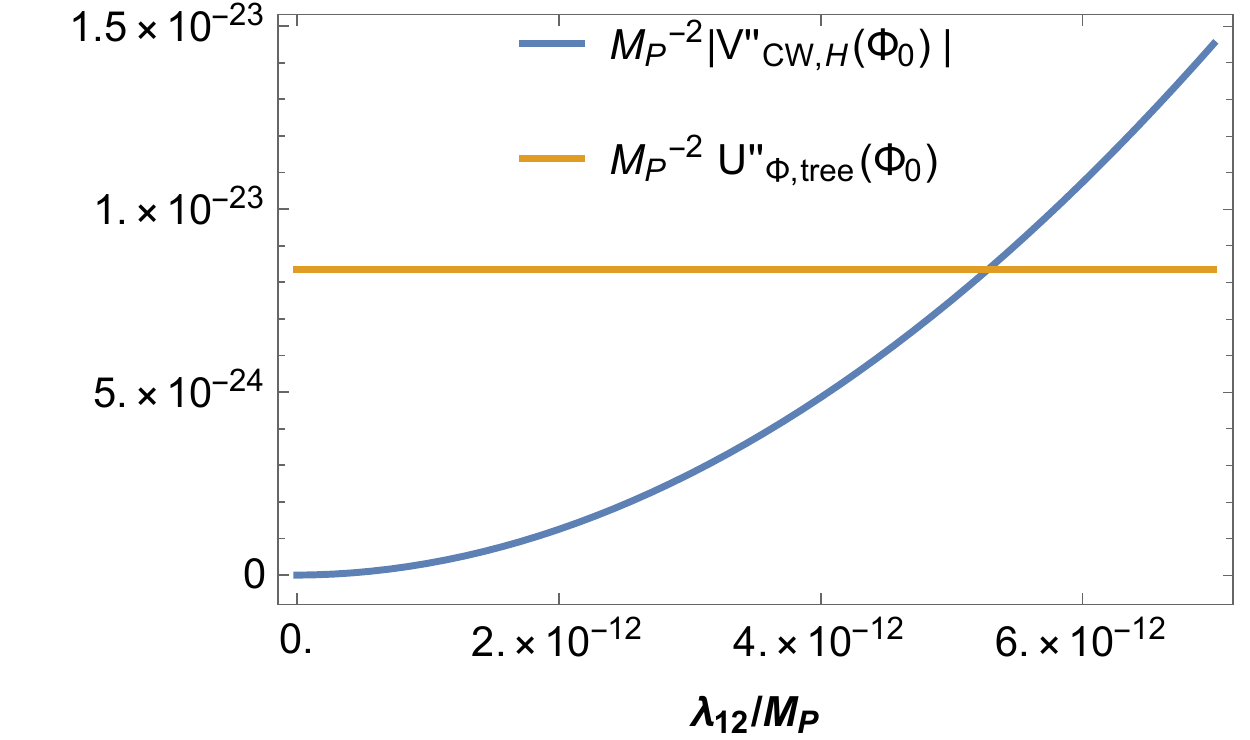}
\end{subfigure}
\vspace{-10pt}
\caption{\it  \raggedright
Allowed range for $y_\chi$ and $\lambda_{12}$ for \text{Model} I inflation from stability. 
The yellow colored line represents the value of tree level potential of \text{Model}~I at $\Phi_0$. The green and blue colored lines indicate the CW correction due to $\chi$ and $H$, respectively. 
}
\label{fig:stability_analysis_plot_for_model_I}
\end{figure}

The upper bound of the value of $\lambda_{12}$  at $\Phi\sim\Phi_{0}$ can be deduced from $\left|V_{{\rm CW},H}^{\prime \prime}\right| 
< V_{\rm tree}^{\prime \prime} (\Phi_{0})$, and it is depicted on the right panel of Fig.~\ref{fig:stability_analysis_plot_for_model_I}. Thus, allowed value of $\lambda_{12}/M_P$ is $<5.283\times 10^{-12}$. 

Similarly,  for $y_\chi$,
\begin{align}
& \left|V_{{\rm CW},\chi}^{\prime \prime} 
\right|= \frac{1}{8 \pi ^2} \left(6 \Phi ^2 y_\chi^4  \ln
\left(\frac{\Phi ^2 y_\chi^2}{\Phi_0^2}\right)-2 \Phi ^2 y_\chi^4 \right)\, .
\label{eq:upper limit of yc model I}
\end{align}
The upper bound on $y_\chi$ around $\Phi\sim\Phi_{0}$ can be obtained from $\left|V_{{\rm CW},\chi}^{\prime \prime} \right|<V_{\rm tree}^{\prime\prime} (\Phi_{0})$ which is exhibited on the left panel of Fig.~\ref{fig:stability_analysis_plot_for_model_I}, and it gives $y_\chi < 4.578\times 10^{-6}$.

\subsection{Stability analysis for sextic inflation}

In this model, inflaton is $\varphi$. Accordingly, the field-depended mass of the fermionic field and Higgs field are respectively 
\begin{eqnarray}
& \widetilde{m}_{\chi}^2 (\varphi) = \left( m_\chi + y_\chi \varphi \right)^2 \,,\\
& \widetilde{m}_{H}^2 (\varphi) = m_H^2 + \lambda_{12} \varphi \,.
\end{eqnarray} 

\begin{figure}[htp]
\centering
\begin{subfigure}{0.45\textwidth}
  \centering
 \includegraphics[width=\linewidth]{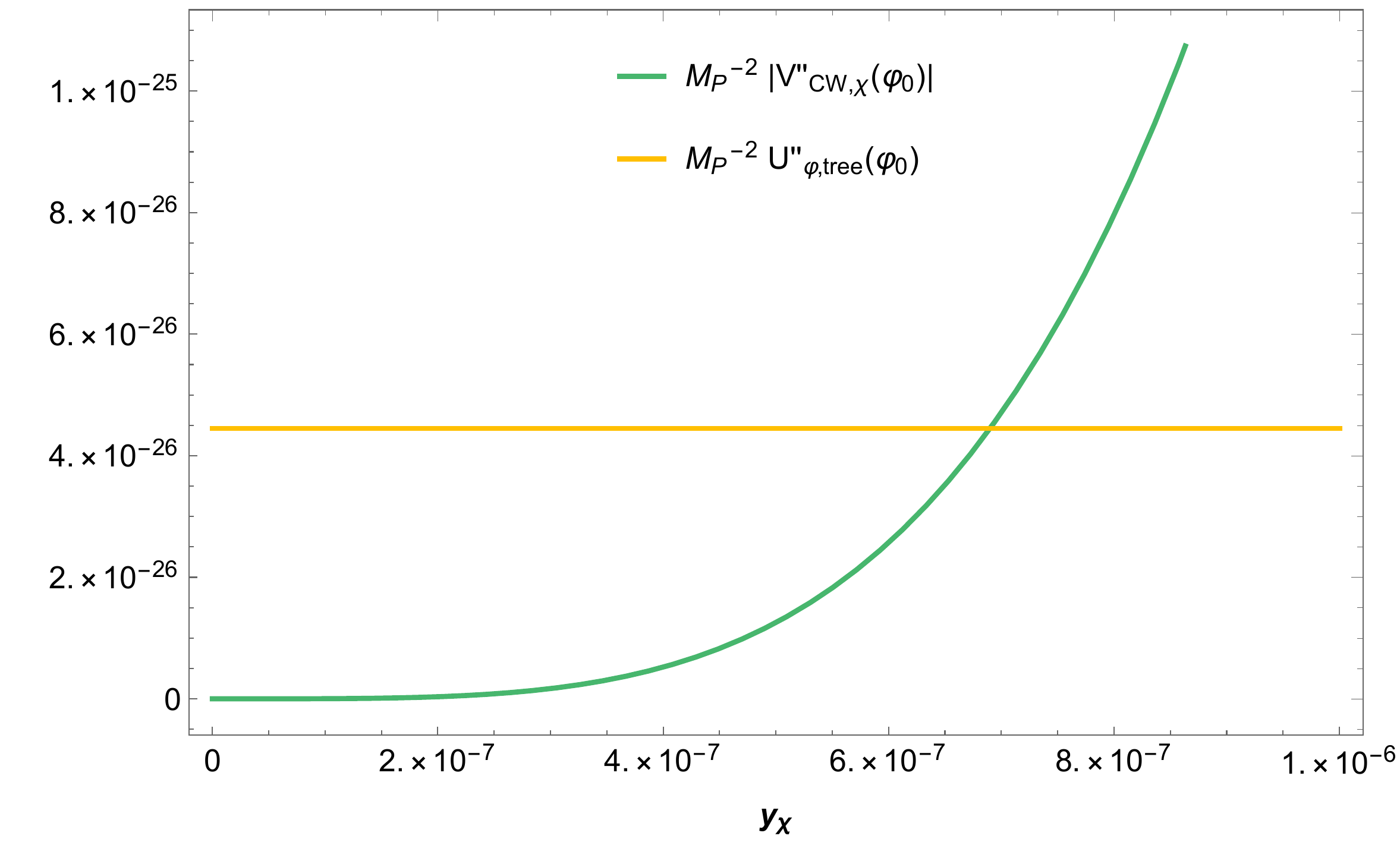}
\end{subfigure}%
\hspace{30pt}
\begin{subfigure}{.45\textwidth}
  \centering
  \includegraphics[width=\linewidth]{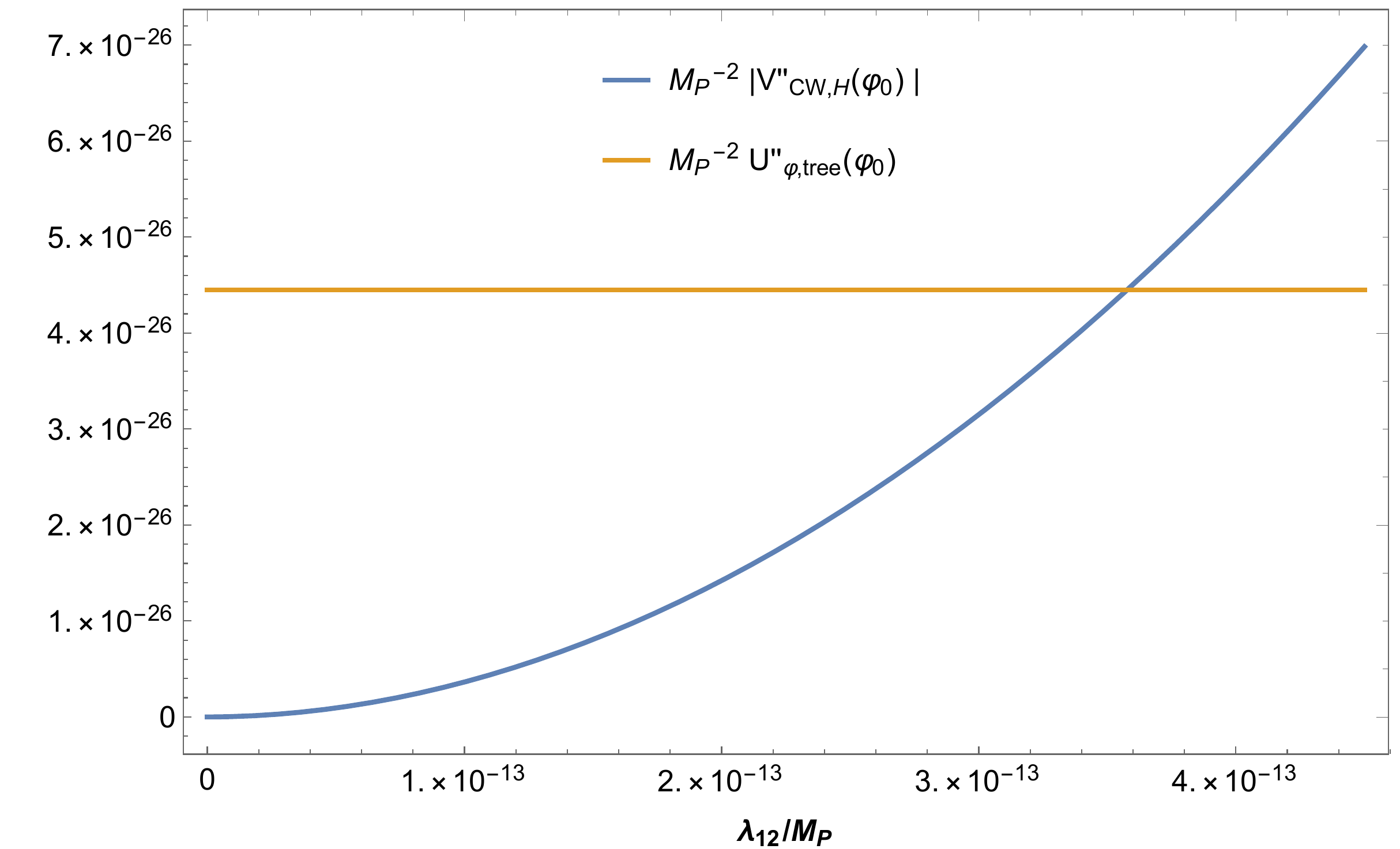}
\end{subfigure}
\vspace{-10pt}
\caption{\it  \raggedright
From the stability analysis of \text{Model} II inflation,
allowed range for $y_\chi$ and $\lambda_{12}$. 
The green and blue colored lines result from CW correction for $\chi$ and $H$, and they are compared with the value of 
tree-level potential of at $\varphi_0$ (yellow colored horizontal line). 
}
\label{fig:stability_analysis_plot_for_model_II}
\end{figure}

From Eq.~\eqref{eq:modified potential of model II}
\begin{align}
& V_{\rm tree}^{\prime\prime} (\varphi_{0}) \equiv U''_\varphi(\varphi_{0}) = \frac{2 q^2}{3 \text{w}}-12 (1-\beta^{II} ) q \varphi_{0}^2+30 (1-\beta^{II} ) \text{w} \varphi_{0}^4 \,,
\end{align}

where $\beta^{II}_1=\beta^{II}_2=\beta^{II}$ (because we have chosen  $\beta^{II}_1=\beta^{II}_2$ in our benchmark value).
Following the steps similar to the ones mentioned in Section~\ref{sec:Stability analysis for linear term inflation}, for $\lambda_{12}$ 
Eq.~\eqref{eq:second derivative test} results in
\begin{align}
& \left|V_{{\rm CW},H}^{\prime \prime}  
\right|= \frac{\lambda_{12}^2 }{8 \pi ^2} \ln
\left(\frac{\lambda_{12} \varphi }{\varphi_{0}^2}\right) \, ,\label{eq:upper limit of l12-coupling model II}
\end{align}
and for $y_\chi$
\begin{align}
& \left| V_{{\rm CW},\chi}^{\prime \prime}
\right|= \frac{1}{8 \pi ^2} \left(6 \varphi ^2 y_\chi^4  \ln
\left(\frac{\varphi ^2 y_\chi^2}{\varphi_{0}^2}\right)-2 \varphi ^2 y_\chi^4 \right). \label{eq:upper limit of yc model II}
\end{align}
In this inflationary case, upper bound on $\lambda_{12}$ and $y_\chi$  around $\varphi\sim\varphi_{0}$ comes from $\left|V_{{\rm CW},H}^{\prime\prime} \right|< V_{\rm tree}^{\prime\prime} (\varphi_{0})$, and $\left|V_{{\rm CW},\chi}^{\prime \prime} \right|<V_{\rm tree}^{\prime\prime} (\varphi_{0})$, respectively. These have been shown in Fig.~\ref{fig:stability_analysis_plot_for_model_II}.  The upper bounds are $y_\chi < 6.9 \times 10^{-7}$, and $\lambda_{12}/M_P< 3.58 \times 10^{-13}$.

\section{Reheating and Dark Matter}
\label{Sec:Reheating}
As soon as the slow roll epoch ends, inflaton quickly drops to the minimum of the potential and starts coherent oscillation about that minimum. If $\Phi_{\rm min}$ (in \text{Model} I) and $\varphi_{\rm min}$ (in \text{Model} II) are the locations of minimum of the inflaton potential respectively, then effective mass of the inflaton in two inflationary models are  
\begin{eqnarray}
\frac{m_{\Phi (\varphi)}}{M_P} = 
\begin{cases}
\left(M_P^{-2}U''_\Phi (\Phi) |_{\Phi=\Phi_{\rm min}} \right)^{1/2} = 6.465 \times 10^{-9} \, \qquad  \text{(for Model I)}   \,,\\
\left(M_P^{-2}U''_\varphi (\varphi) |_{\varphi=\varphi_{\rm min}} \right)^{1/2} =  1.705\times 10^{-9}\,  \qquad  \text{(for Model II)} \,. 
\end{cases}
\end{eqnarray} 

This oscillating field acts as a non-relativistic fluid without any pressure when averaged over a number of coherent oscillations. The energy density of this inflaton decreases due to two reasons - Hubble expansion and decay to 
relativistic {SM}~Higgs particle $h$ and {DM}~particle $\chi$ following the Lagrangian density of Eq.~\eqref{eq:reheating lagrangian for modelI} and Eq.~\eqref{eq:reheating lagrangian for modelII}. The decay width of inflaton to $h$ and $\chi$ are
\begin{align}
	\Gamma_{\Phi (\varphi) \to {h h}}  \simeq \frac{\lambda_{12}^2}{8\pi\, m_{\Phi(\varphi)}}\,, 
 \qquad 
 \Gamma_{\Phi(\varphi) \to \chi\chi}  \simeq \frac{y_\chi^2\, m_{\Phi (\varphi)}}{8\pi}\,. %
\end{align}	
To satisfy present-day relic density of photons and baryons, 
we are considering 
$\Gamma_{\Phi(\varphi)  \to {h h}}  > \Gamma_{\Phi(\varphi)  \to \chi\chi}  $ 
such that total decay width of inflaton $\Gamma =\Gamma_{\Phi(\varphi)  \to \chi\chi} + \Gamma_{\Phi(\varphi)  \to {h h}} \simeq \Gamma_{\Phi(\varphi)  \to {h h}}$. Hence,  
\begin{eqnarray}
\Gamma = 
\begin{cases}
6.15 \times 10^6 \, \frac{\lambda_{12}^2}{ M_P} \qquad  \text{(for Model I)}  \,,\\
2.33 \times 10^7 \, \frac{\lambda_{12}^2}{ M_P}                    \qquad  \text{(for Model II)}  \,. %
\end{cases} 
\end{eqnarray} 

Now, the branching ratio for the production of $\chi$ is 
\begin{align}
\text{Br} &= \frac{\Gamma_{\Phi (\varphi) \to \chi\chi} }{\Gamma_{\Phi (\varphi) \to \chi\chi} + \Gamma_{\Phi (\varphi) \to {h h}} } \simeq  \frac{\Gamma_{\Phi (\varphi) \to \chi\chi} }{ \Gamma_{\Phi (\varphi) \to {h h}} }   = m_{\Phi (\varphi)}^2  \left( \frac{y_\chi}{\lambda_{12}}\right)^2\\
&= 
\begin{cases}
4.18 \times 10^{-17} \left(\frac{y_\chi}{\lambda_{12}} \right)^2 M_P^2  \qquad  \text{(for Model I)}  \,,\\
  2.91 \times 10^{-18}    \left(\frac{y_\chi}{\lambda_{12}} \right)^2  M_P^2                 \qquad  \text{(for Model II)}  \, .
\end{cases} 
\end{align}

These produced particles cause the development of the local-thermal relativistic fluid of
the universe and consequently, raise the temperature of the universe. At the beginning of reheating, 
due to the small value of couplings to inflaton, $\Gamma< {\cal H} (\ss)$, where ${\cal H}\equiv{\cal H} (\ss)$ is the Hubble parameter  and $\ss$ is the cosmological scale factor. 
Meanwhile, ${\cal H}$ continues to decrease. 
At the moment when ${\cal H}$ becomes $\sim \Gamma$, the temperature of the universe is called as reheating temperature, $T_{rh}$, and it is can be computed as~\cite{Bernal:2021qrl}
\begin{align}\label{eq:definition of reheating temperature}
T_{rh} 
= \sqrt{\frac{2}{\pi}} \left(\frac{10}{g_{\star}}\right)^{1/4} \sqrt{M_P} \sqrt{\Gamma}\,
=
\begin{cases}
1095.07 \, \lambda_{12} \qquad  \text{(for Model I)}  \,,\\
2132.09 \lambda_{12}      \qquad  \text{(for Model II)} \, .
\end{cases} 
\end{align}

We have assumed $g_{\star} = 106.75$.
At temperature below $T_{rh}$, the universe behaves as if it is dominated by relativistic particles~\cite{Kolb:2003ke}.
Additionally, we have assumed here that the process of particle production from inflaton is instantaneous~\cite{Chung:1998rq}. In general, reheating is not an instantaneous process. The maximum temperature of the universe during the whole process of reheating may be many orders greater than $T_{rh}$
and it can be estimated as%
~\cite{
Chung:1998rq}
\begin{align} \label{eq:TMAX}
T_{max} = 
\Gamma^{1/4} \left( \frac{60}{g_{\star} \pi^2} \right)^{1/4} \left( \frac{3}{8}\right)^{2/5} {\cal H}_I^{1/4} M_P^{1/2} \,,
\end{align}
where ${\cal H}_I$ is the value of the Hubble parameter at the beginning of reheating when no particle, including the {DM}, is produced. This can be taken as 
\begin{align}\label{eq:HI}
{\cal H}_I 
\simeq 
\begin{cases}
\sqrt{\frac{U_\Phi (\Phi_{0} )}{3 M_P^2}}
=
3.23 \times 10^{-10} M_P \qquad  \text{(for Model I)} %
\,,\\ 
\sqrt{\frac{U_\varphi (\varphi_{0} )}{3 M_P^2}}= 1.206 \times 10^{-10} M_P        \qquad  \text{(for Model II)} %
\, .
\end{cases} 
\end{align}
The Eq.~\eqref{eq:definition of reheating temperature}
with $T_{rh} \gtrsim 4 \text{MeV}$ puts down the lower limit on $\lambda_{12}$ 
\begin{eqnarray}\label{eq:lower limit of coupling}
\frac{\lambda_{12}}{M_P}\gtrsim 
\begin{cases}
1.52 \times 10^{-24} \qquad  \text{(for Model I)}  \,,\\
7.82 \times 10^{-25} \qquad  \text{(for Model II)} \, .
\end{cases} 
\end{eqnarray} 

%
From Eq.~\eqref{eq:TMAX}, we can write
\begin{eqnarray}\label{Eq:TmaxTrh-ratio}
\frac{T_{max}}{T_{rh}} = \left(\frac{3}{8}\right)^{2/5}  \left( \frac{{\cal H}_I}{{\cal H}(T_{rh})} \right)^{1/4}\,,
\end{eqnarray}
where
\begin{eqnarray}
{\cal H}(T_{rh}) = \frac{\pi}{3 M_P} \sqrt{\frac{g_{\star}}{10}}  T_{rh}^2  \,.
\end{eqnarray} 

\begin{figure}[htp]
\centering
\begin{subfigure}{0.45\textwidth}
  \centering
 \includegraphics[width=\linewidth]{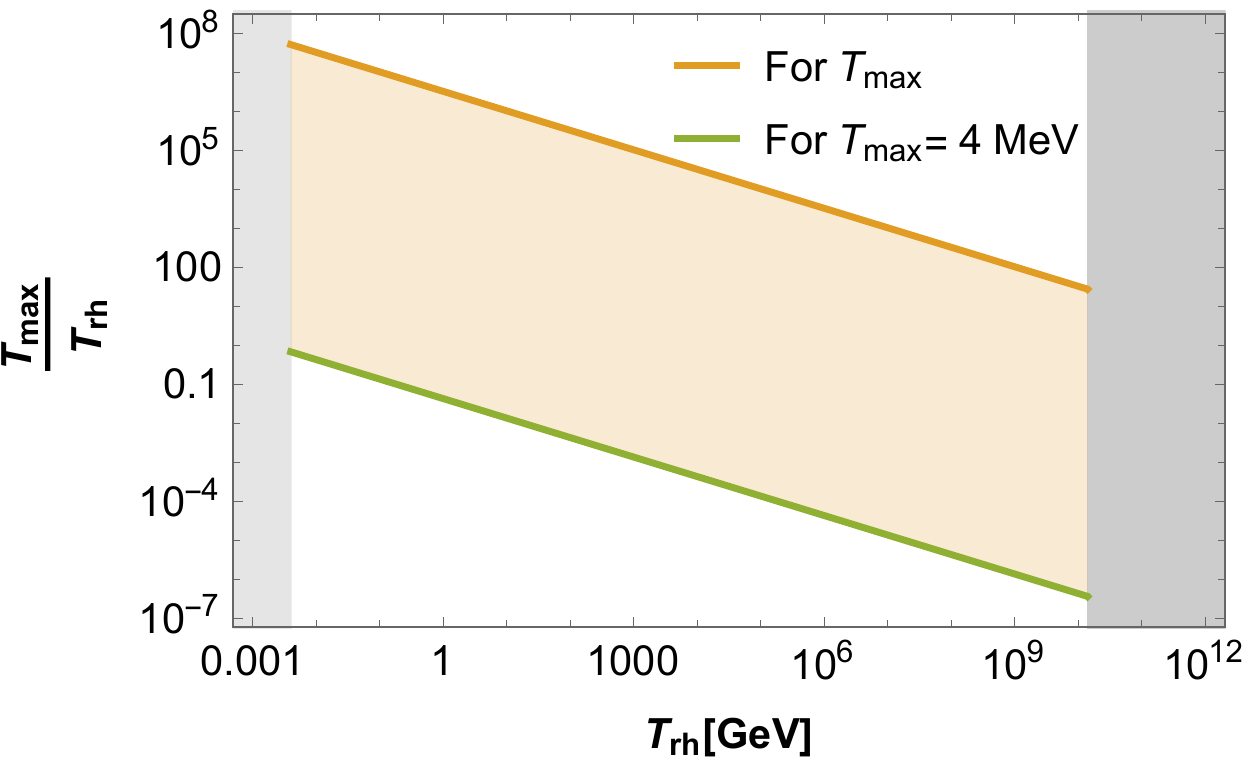}
\end{subfigure}%
\hspace{30pt}
\begin{subfigure}{.45\textwidth}
  \centering
  \includegraphics[width=\linewidth]{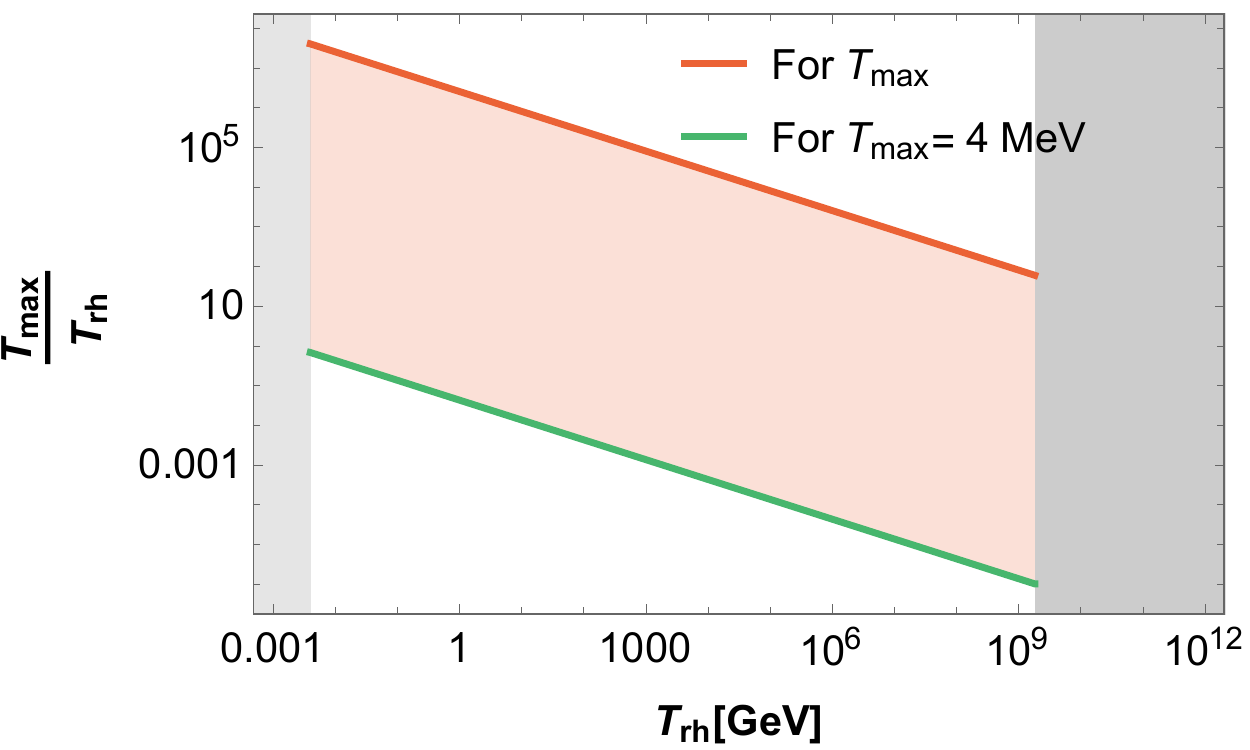}
\end{subfigure}
\vspace{-10pt}
\caption{\it  \raggedright
Allowed range (colored region) for $T_{max}/T_{rh}$: left panel is for \text{Model}~I inflation
, where right panel is for the \text{Model}~II.
The green color line points to $T_{max}/T_{rh}$ when $T_{max}=4 \text{MeV}$. 
The gray colored area indicates the lower ($T_{rh} \nless 4 \text{MeV}$) and upper bound on $T_{rh}$ obtained from the stability analysis (see Eq.~\eqref{eq:upper limit of l12-coupling model I} and Eq.~\eqref{eq:upper limit of l12-coupling model II}).
}
\label{fig:allowed range for Tmax/Trh}
\end{figure}

The allowed ranges for $T_{max}/T_{rh}$ for two inflationary models are shown in Fig.~\ref{fig:allowed range for Tmax/Trh}. The upper limit for the allowed region comes from Eq.~\eqref{Eq:TmaxTrh-ratio} and the lower limit from the fact that $T_{rh}\gtrsim4\text{MeV}$ which is needed for successful Big Bang nucleosynthesis (BBN)
~\cite{Giudice:2000ex}.

\subsection{Dark Matter Production and Relic Density}
In this subsection, we estimate, following Ref.~\cite{Bernal:2021qrl}, the amount of {DM}~produced during reheating and compared it with {DM}~relic density of the present-day universe. The Boltzmann equation for the 
evolution of 
{DM}~number density, $n_\chi$, of {DM}~particles 
is -
	\begin{equation}
	    \frac{\text{d} n_\chi}{\text{d} t} + 3{\cal H}\,n_\chi  = \gamma\,,
	\end{equation}
where $t$ is the physical time, $\gamma$ is the rate of production of {DM}~per unit volume. 
Then the evolution equation of comoving number density, $N_\chi= n_\chi \ss^3$ ($\ss(t)$ is the cosmological scale factor, as mentioned earlier), 
of {DM}~particles
\begin{eqnarray}\label{eq:Boltzaman equation for comoving number density}%
\frac{\text{d} N_\chi}{\text{d} t} = \ss^3 \gamma  \,.
\end{eqnarray} 
%
While the temperature, $T$ of the universe is $T_{max} > T>T_{rh} $, the energy density of the universe is dominated by inflaton and the first Friedman equation leads to 
~\cite{Bernal:2021qrl}
\begin{eqnarray}\label{eq:Hubble parameter during reheating}
{\cal H} = \frac{\pi}{3} \sqrt{\frac{g_{\star}}{10}} \frac{T^4}{M_P\, T_{rh}^2}\,.
\end{eqnarray} 

Therefore, energy density of inflaton
\begin{eqnarray}
\rho_{\Phi(\varphi)} = \frac{ \pi^2  g_{\star} }{30 } \frac{T^8}{ T_{rh}^4} \,.
\end{eqnarray}

Since, during reheating, $\rho_\Phi$ behaves as a non-relativistic fluids, $\rho_{\Phi(\varphi)} \propto \ss^{-3}$, 
the scale factor behaves as
\begin{eqnarray}\label{eq:rheating sclae factor}
\ss\propto T^{-8/3}\,.
\end{eqnarray} 

Using Eq.~\eqref{eq:Hubble parameter during reheating} and \eqref{eq:rheating sclae factor} in Eq.~\eqref{eq:Boltzaman equation for comoving number density} we obtain
\begin{eqnarray}\label{eq:dNchidT}
\frac{\text{d} N_\chi}{\text{d} T} = - \frac{8 M_P }{\pi} \left(\frac{10}{  g_{\star}}\right)^{1/2} \frac{T_{rh}^{10}}{T^{13}} 
\, \ss^3 (T_{rh}) \, \gamma  \,.
\end{eqnarray} 

{DM}~Yield, $Y_\chi$ is defined as the ratio of the number density of {DM}~to the entropy density of photons, i.e., 
$
Y_\chi = \frac{n_\chi(T)}{s(T)}\, ,
$
where entropy density $s(T)=\frac{2 \pi^2}{45} g_{\star, s} T^3 $ and $g_{\star, s}$ is the effective number of  degrees of freedom of the constituents of the relativistic fluid. 
If we assume that there is no entropy generation in any cosmological process, after reheating epoch, 
then the evolution of $Y_\chi$ can be expressed as
	\begin{equation} \label{eq:evolution of Yield}
	    \frac{\text{d}Y_\chi}{\text{d} T} = - \frac{135}{2\pi^3\, g_{\star, s}} \sqrt{\frac{10}{g_{\star}}}\, \frac{M_P}{T^6}\, \gamma\,.
	\end{equation}

We are assuming that
the {DM}~particles, produced during reheating, were never in thermal equilibrium with the relativistic fluid of the universe. 
Those DM particles contribute to the cold dark matter ({CDM}) density of the present universe.
Thus, following Table~\ref{Table:data about CDM}, present-day {CDM}~yield~\cite{Bernal:2021qrl} is 

\begin{eqnarray}\label{eq:present day CDM yield}
 Y_{{\rm CDM},0}  = \frac{4.3. \times 10^{-10}}{m_{\chi}} \,,
\end{eqnarray} 
where $m_{\chi}$ is expressed in $\text{GeV}$. Now, the amount of {DM}~produced during reheating through decay or via scattering in both \text{Model} I and \text{Model} II, has been estimated and compared with $ Y_{{\rm CDM},0}$ in the following part of this subsection.

\begin{table}[ht]
\begin{center}
\caption{ \it Data about CDM ($h_{{CMB}}\approx0.674$)} \label{Table:data about CDM}
\begin{tabular}{ |c| c| c| }
\hline
\multicolumn{1}{|c|}{$\Omega_{\rm CDM}$} & \multicolumn{1}{c|}{ $ 0.120 \, h_{{CMB}}^{-2} $} & \multirow{3}{*}{\cite{ParticleDataGroup:2020ssz}} \\
\cline{1-2}
\multicolumn{1}{|c|}{$\rho_c$} & \multicolumn{1}{c|}{$1.878 \times 10^{-29} \, h_{{CMB}}^2 \, \text{g} \text{cm}^{-3}$} & \\
\cline{1-2}
\multicolumn{1}{|c|}{$s_0$} & \multicolumn{1}{c|}{$2891.2\,  (T/2.7255 \text{K})^3 \, \text{cm}^{-3}$} & \\
\hline 
\end{tabular}
\end{center}
\end{table}%
%
%
%

\subsubsection{Inflaton decay}
%
If {DM}~particles are generated from the inflaton decay
%
\begin{eqnarray}
\gamma = 2 \text{Br} \, \Gamma \,  \frac{\rho_{\Phi(\varphi)}}{m_{\Phi(\varphi)}} \,.
\end{eqnarray} 
Substituting this in Eq.~\eqref{eq:evolution of Yield}, the {DM}~yield from the decay of inflaton,
%
\begin{align}
Y_{\chi,0}%
&\simeq \frac{3}{\pi} \frac{g_{\star}}{g_{\star, s}} \sqrt{\frac{10}{g_{\star}}} \frac{M_P \, \Gamma}{m_{\Phi(\varphi)}\, T_{rh}} \text{Br}%
= \frac{3}{\pi} \frac{g_{\star}}{g_{\star, s}} \sqrt{\frac{10}{g_{\star}}} \frac{M_P }{ T_{rh}}  \frac{(y_\chi)^2}{8 \pi} \\
 &= 1.163\times 10^{-2} M_P \frac{y_\chi^2 }{T_{rh}}\,. \label{ychi0-decay-total}
\end{align}

%
Here, we assume $g_{\star, s}=g_{\star}$. Equating Eq.~\eqref{ychi0-decay-total} with Eq.~\eqref{eq:present day CDM yield}, we get 
the condition to generate the complete {CDM}~energy density -%
\begin{eqnarray}\label{eq:eq to plot 2}
T_{rh}  \simeq  
6.49 \times 10^{25}y_\chi^2 m_{\chi} \,.
\end{eqnarray} 

\begin{figure}[htp]
\centering
\begin{subfigure}{0.45\textwidth}
  \centering
 \includegraphics[width=\linewidth]{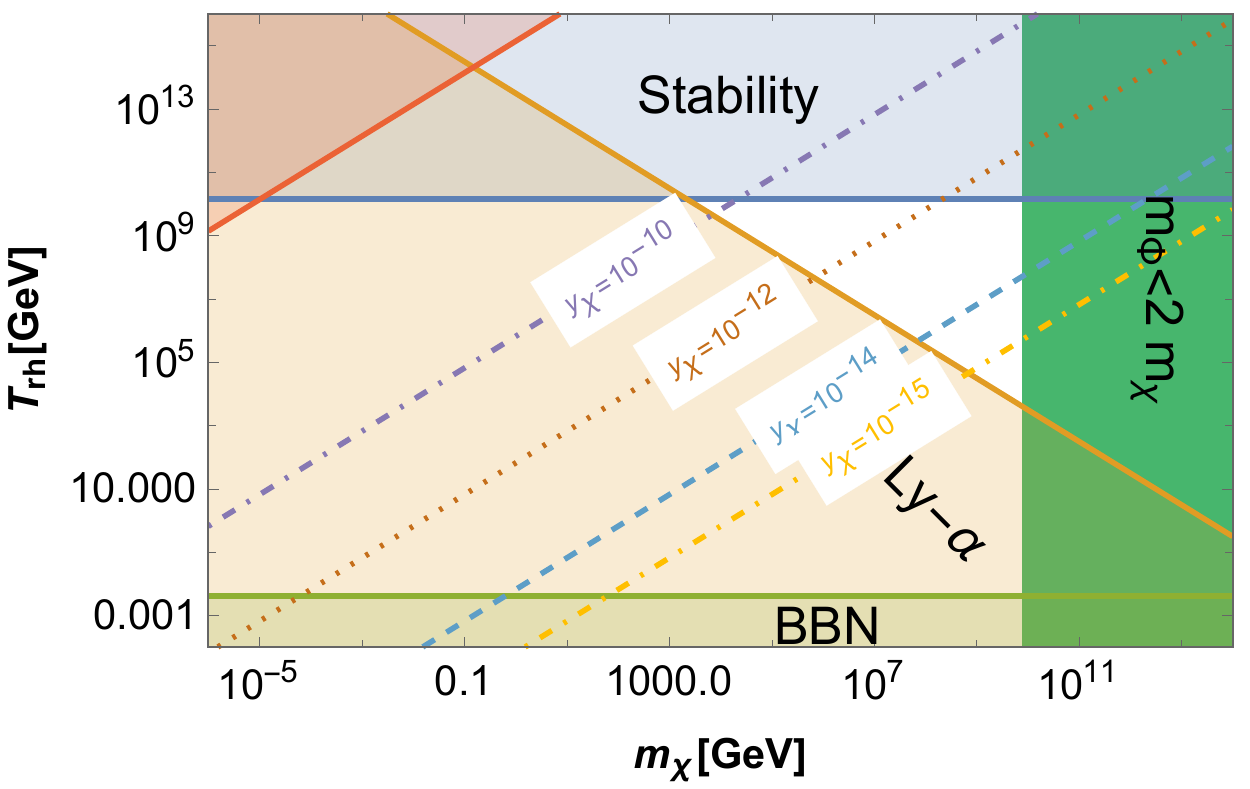}
\end{subfigure}%
\hspace{30pt}
\begin{subfigure}{.45\textwidth}
  \centering
  \includegraphics[width=\linewidth]{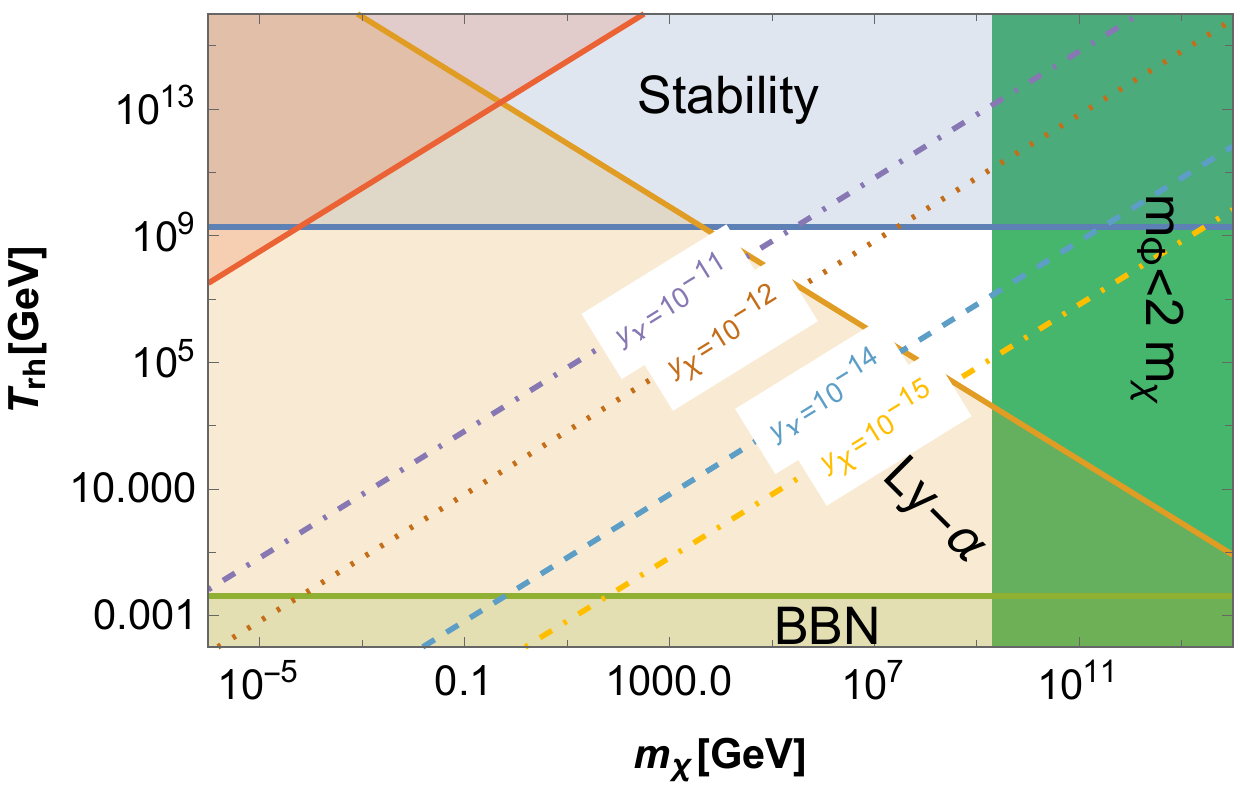}
\end{subfigure}
\vspace{-10pt}
\caption[Caption for LOF]{\it  \raggedright
The allowed region (unshaded) for the Yukawa-like coupling $y_\chi$ to produce the complete {CDM}~of the present universe: left panel is for \text{Model} I inflation and right for \text{Model} II inflation. The constraints (colored regions) are from (a) BBN (light green colored region): $T_{rh}> 4 \text{MeV}$, 
(b) from stability analysis (blue colored region): $T_{rh} \simeq 1.388 \times 10^{10} \text{GeV}$~\text{(for Model I)} or $T_{rh} \simeq 1.83 \times 10^{9} \text{GeV}$~\text{(for Model II)} from the upper bound of $\lambda_{12}$ from Eq.~\eqref{eq:upper limit of l12-coupling model I} or Eq.~\eqref{eq:upper limit of l12-coupling model II}, (c) stability (red-colored region): from the upper bound of $y_\chi$ from Eq.~\eqref{eq:upper limit of yc model I} or Eq.~\eqref{eq:upper limit of yc model II}, (d) (deep green region): $m_{\chi}$ must be $<m_\Phi/2$ (\text{Model} I) or $<m_\varphi/2$ (\text{Model} II), (e) (light peach-colored region):~Ly-$\alpha$
:~$T_{rh} \gtrsim (2  m_\Phi)/m_{\chi} $ or $T_{rh} \gtrsim (2  m_\varphi)/m_{\chi}$~\cite{Bernal:2021qrl}. 
}
\label{fig:allowed range of ychi}
\end{figure}

Fig.~\ref{fig:allowed range of ychi} depicts the allowed range of the coupling $y_\chi$ from Eq.~\eqref{eq:eq to plot 2}, to generate the complete {CDM}~density of the contemporary universe only via the decay channel of inflaton.  
From this figure, we can deduce that the allowed range for $y_\chi$ and $m_{\chi}$ to construct the {CDM}~density of the universe is 
$10^{-10} \gtrsim y_\chi \gtrsim 10^{-15}$ (for $2.5\times 10^3 \, \text{GeV} \lesssim \, m_{\chi} \, \lesssim 8.1\times 10^{9} \, \text{GeV}$ in \text{Model} I) and  $10^{-11} \gtrsim y_\chi \gtrsim 10^{-15}$ (for $8.4\times 10^3 \, \text{GeV} \lesssim \, m_{\chi} \, \lesssim 2\times 10^{9} \, \text{GeV}$ in \text{Model} II).

\subsubsection{{DM}~production from scattering channel}
\label{Inflaton-Scattering}
\begin{figure}[htp]%
\centering
\begin{subfigure}{0.45\textwidth}
  \centering
 \includegraphics[width=\linewidth]{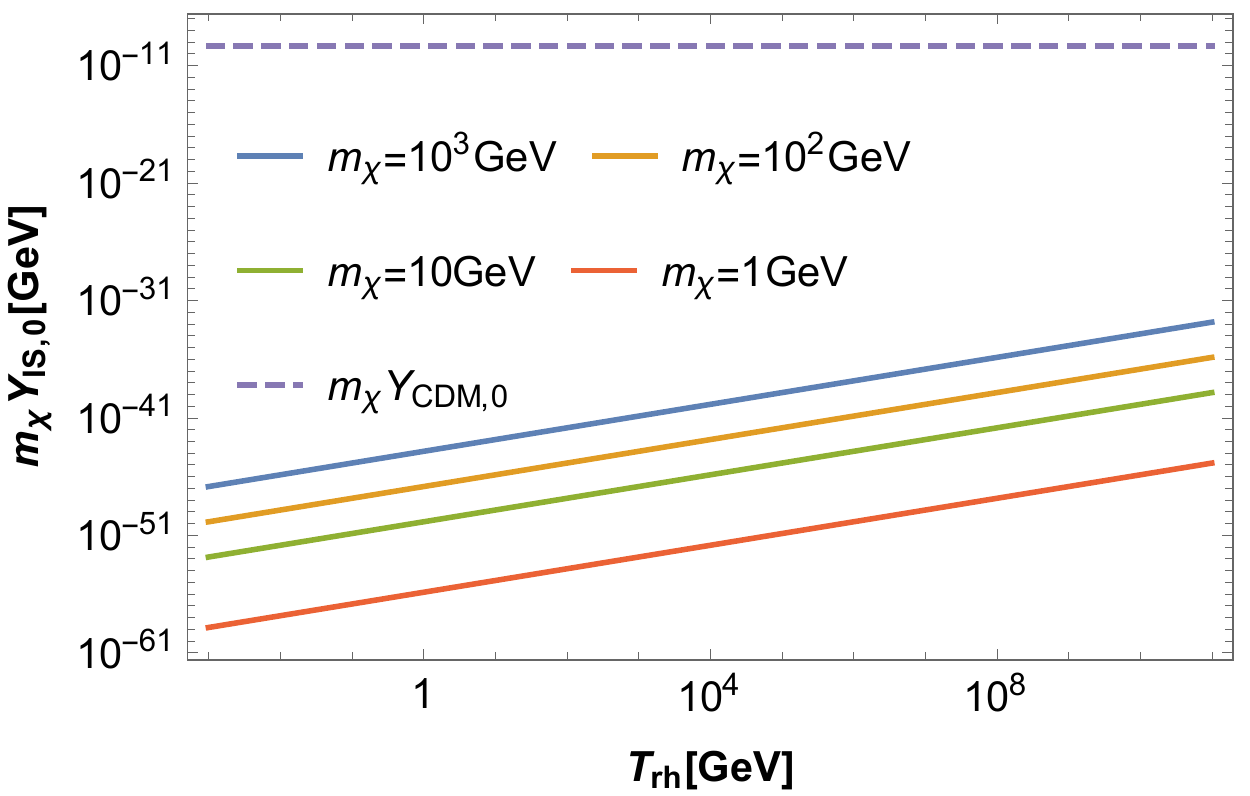}
\end{subfigure}%
\hspace{30pt}
\begin{subfigure}{.45\textwidth}
  \centering
  \includegraphics[width=\linewidth]{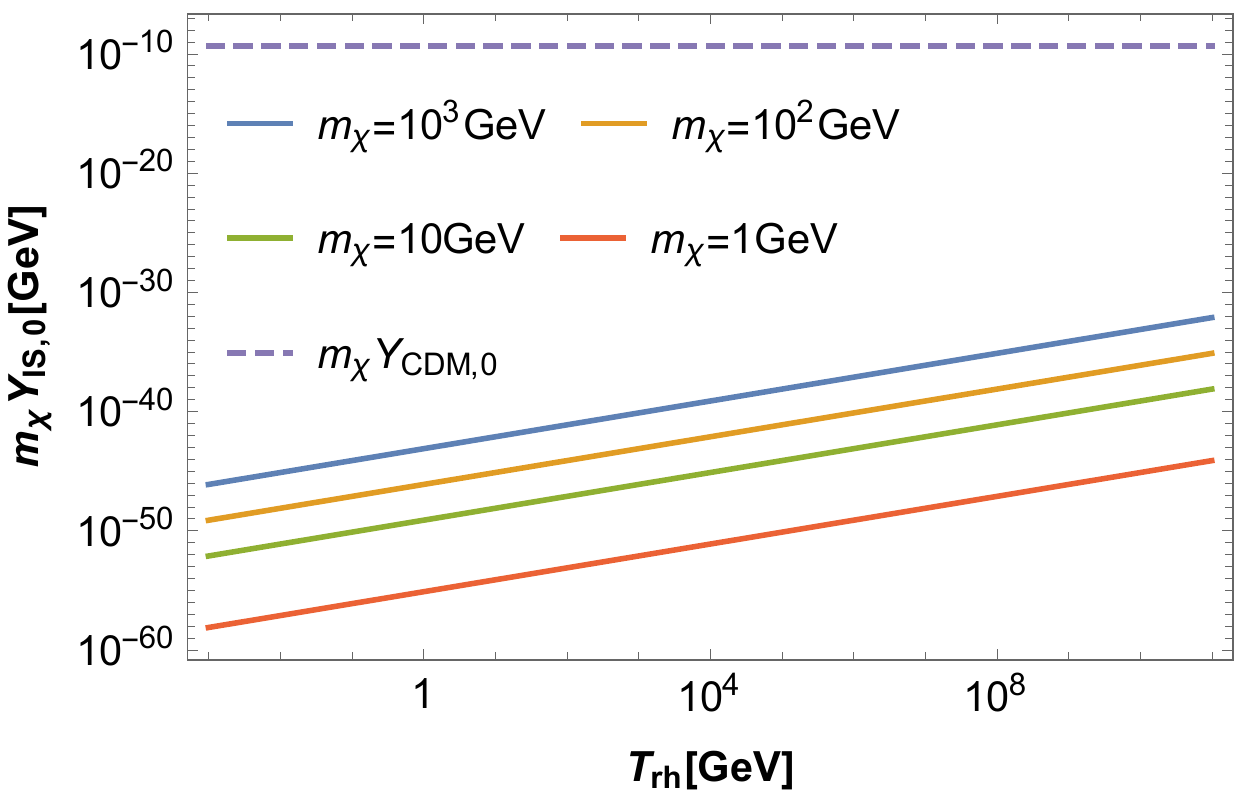}
\end{subfigure}
\vspace{-10pt}
\caption{\it  \raggedright
$m_{\chi} \times$yield of {DM}~generated from the 2-to-2 scattering with graviton as mediator for different values of $m_{\chi}$. The left panel shows the result for \text{Model} I and the right panel for \text{Model} II inflation.
}
\label{fig:mc YIS0 - from inflaton scattering via graviton}
\end{figure}

In this work, we consider the 2-to-2 scattering processes which contribute significantly in {DM}~production, as mentioned in~\cite{Bernal:2021qrl}.
When graviton acts as the mediator for the production of {DM}~particles from non-relativistic inflaton via 2-to-2 scattering, then 
the {DM}~yield~\cite{Bernal:2021qrl} 
	\begin{align} \label{eq:yield-DM-scattering-inflaon-graviton}
	    Y_{IS,0} &\simeq \frac{g_{\star}^2}{81920g_{\star, s}} \sqrt{\frac{10}{g_{\star}}} \left(\frac{T_{rh}}{M_P}\right)^3 \left[\left(\frac{T_{max}}{T_{rh}}\right)^4 - 1 \right] \frac{m_{\chi}^2}{m_{\Phi(\varphi)}^2} \left(1-\frac{m_{\chi}^2}{m_{\Phi(\varphi)}^2}\right)^{3/2}\,.
    \end{align}
In Fig.~\ref{fig:mc YIS0 - from inflaton scattering via graviton}, $Y_{IS,0}$ {(actually $m_{\chi} Y_{IS,0}$ with $m_{\chi} Y_{{\rm CDM},0}$)}  is compared with $Y_{{\rm CDM},0}$ for different $m_{\chi}$ as a function of $T_{rh}$. Hence, it is shown there that the yield  of {DM}~ produced via scattering (Eq.~\eqref{eq:yield-DM-scattering-inflaon-graviton}) is not significant compared to the present {CDM}~density.

%
{DM}~particles can also be produced from the scattering of {SM}~particles via graviton mediation. In that case,
\begin{equation}
	\gamma = \alpha\, \frac{T^8}{M_P^4}\,,
\end{equation}
where $\alpha\simeq 1.1 \times 10^{-3}$. 
Due to the presence of $M_P^4$ in the denominator, it is expected that the production of {DM}~through this process is less compared to previous ones and thus, we neglect. 


When inflaton acts as mediator for the production of {DM}~from  2-to-2 scattering of {SM}~particles,
production of {DM}~(yield) only through that channel results in 

	\begin{equation} 
	Y_{{SM} i,0} \simeq \frac{135\,  y^2_{\chi}\, \lambda_{12}^2}{4 \pi^8\, g_{\star, s}} \sqrt{\frac{10}{g_\star}}\, \frac{M_P\, T_{rh}}{m_{\Phi(\varphi)}^4}\, , \qquad \text{ for } T_{rh} \ll m_{\Phi(\varphi)}, T_{rh} > T \,.
	\end{equation}

$Y_{{SM} i,0}\sim 10^{-60}$ ($\sim 10^{-62}$) for $T_{rh}\sim 10^{5} \text{GeV} \simeq 10^{-5} m_\Phi$ ($m_\varphi$) for $g_{\star}=g_{\star, s}=106.75$, $\lambda_{12}\sim 10^{-12}$ ($10^{-13}$) and $y_\chi \sim 10^{-6}$ ($10^{-7}$). 
Therefore, the {DM}~produced from 2-to-2 scattering during reheating is 
insignificant in comparison to total {CDM}~density of the universe.

\section{Conclusions and Discussion}
\label{Sec:Conclusions and Discussion}
We investigated a simple possibility of a scalar inflaton and a non-thermal fermionic particle that originated during the reheating epoch and acted as the CDM. Satisfying the correct relic density of {DM}~and other {CMB}~bounds, we discovered the following features of our analysis:
\begin{itemize}
\item We investigated two polynomial potential models for slow roll single field cosmic inflation. Each of these models features an inflection point. Moreover, due to the presence of a term corresponding to the linear power of inflaton (see Eq.~\eqref{eq:inflation potential of model I}), the potential of \text{Model}~I is not symmetric about the origin. In contrast, the potential of Model II (Eq.~\eqref{eq:inflation potential of model II}) is symmetric under the transformation of $\varphi\to -\varphi$.

\item We computed the coefficients of the potentials of both models satisfying the current {CMB}~bounds and under the assumption of near-inflection point inflationary scenario. We also found $n_s\sim 0.96$, $r\sim 10^{-12}, \alpha_s\sim 10^{-3}$, and $\beta_s\sim 10^{-8}$  (see Table~\ref{Tab:Model I benchmark values} and Table~\ref{Tab:Model II benchmark values}). 

\item We assumed that inflaton decays to {SM}~Higgs ($H$) together with {DM}~$(\chi)$. 
From stability analysis of the inflation-potential in Fig.~\ref{fig:stability_analysis_plot_for_model_I} and Fig.~\ref{fig:stability_analysis_plot_for_model_II}, we deduced that the upper bounds of the couplings for two decay channels are $\lambda_{12}/M_P \lesssim {\cal O}(10^{-12})$ and $y_\chi \lesssim {\cal O}(10^{-6})$. The former upper bound defines the highest permissible value of $T_{rh}$.

\item We studied the formation of non-thermal vector-like fermionic {DM}~particles, during reheating from the inflaton decay. The rate of {DM}~creation through this decay is temperature dependent; when the temperature of the universe's relativistic fluid increases during reheating, the rate of {DM}~generation reduces (Eq.~\eqref{eq:evolution of Yield}). Fig.~\ref{fig:allowed range for Tmax/Trh} depicts the permissible range for the ratio of the highest temperature $T_{max}$ to the reheating temperature $T_{rh}$ during that period, $T_{max}/T_{rh}$. For $T_{rh}=4\text{MeV}$, the ratio might reach ${\cal O}(10^{7})$. The permitted range of $T_{max}/T_{rh}$ is determined by the inflection point (see Eq.~\eqref{eq:TMAX} and Eq.~\eqref{eq:HI}). Because we chose the {CMB}~scale around the inflection point, the inflection point determines the {CMB}~observables, such as $n_s$ and $r$ on one hand, and controls the production regimes (via T$_{max}$) of {DM}~and consequently {DM}~relic on the other hand.

\item  Fig.~\ref{fig:allowed range of ychi} depicts the allowed region in $T_{rh}-m_{\chi}$ space for two models of potential we have considered and
the
constraints on that space are coming from bound on $T_{rh}$ from {BBN}, radiative stability analysis of the potential for slow roll inflation, Ly-$\alpha$ bound, and the maximum possible value of $m_{\chi}$ for the effective mass of the inflaton. From this figure we can conclude that $\chi$ produced only through the decay of inflaton may explain the total  density of {CDM}~of the current universe if $10^{-10} \gtrsim y_\chi \gtrsim 10^{-15}$ (for $2.5\times 10^3 \, \text{GeV} \lesssim \, m_{\chi} \, \lesssim 8.1\times 10^{9} \, \text{GeV}$ in \text{Model} I) and  $10^{-11} \gtrsim y_\chi \gtrsim 10^{-15}$ (for $8.4\times 10^3 \, \text{GeV} \lesssim \, m_{\chi} \, \lesssim 2\times 10^{9} \, \text{GeV}$ in \text{Model} II).
\\

\item  $\chi$ can also be produced from 2-to-2 scattering of either {SM}~particles or inflatons. Among all those scattering processes, the promising one is – from the scattering of inflaton with graviton as the mediator.~
In Fig.~\ref{fig:mc YIS0 - from inflaton scattering via graviton} we showed that $Y_\chi$ produced through 2-to-2 scattering of inflaton with graviton as mediator, is {more} than the {DM}~production via other scattering channels, and it is $ Y_{IS,0}\sim  {\cal O}( 10^{-36}$) for $T_{rh}=10^8~\text{GeV}, m_{\chi}= 10^3~\text{GeV}$.
But, $ Y_{IS,0}$ produced through this channel is much less than $Y_{{\rm CDM},0}$ and thus $\chi$ produced through 2-to-2 scattering channels can  contribute only a negligible fraction of $Y_{{\rm CDM},0}$.
\end{itemize}

In conclusion, we consider two members of the beyond the standard model~physics - inflaton and the non-thermal {DM}, to connect the {CMB}~data and the {DM}~mystery. This work can be further extended to study the formation of Primordial Black Holes for inflection point inflationary scenario, non-Gaussianities in the {CMB}~spectrum, and generation of Gravitational Waves which can be tested from future {CMB}~experiments.

\medskip

\section*{Acknowledgement}
Shiladitya Porey wants to thank Professor Norma Susana Mankoč Borštnik, Professor Maxim Khlopov, Professor 
Astri Kleppe, and the organizers of the Bled 25th Workshop. Work of Shiladitya Porey is funded by RSF Grant 19-42-02004. Supratik Pal thanks Department of Science and Technology, Govt. of India for partial support through Grant No. NMICPS/006/MD/2020-21.

\end{document}